\documentclass[%
prx,
reprint,
superscriptaddress,
nofootinbib,
 amsmath,amssymb,
 aps,
floatfix,
]{revtex4-2}

\usepackage[columnwise]{lineno}

\let\oldalign\align
\let\oldendalign\endalign
\renewenvironment{align}{%
    \linenomathNonumbers\oldalign%
    }{%
    \oldendalign\endlinenomath%
    }

\let\oldequation\equation
\let\oldendequation\endequation
\renewenvironment{equation}{%
    \linenomathNonumbers\oldequation
    }{%
    \oldendequation\endlinenomath%
    }

\usepackage{graphicx}
\usepackage{dcolumn}
\usepackage{bm}
\usepackage{bbold}
\usepackage{bbm}

\usepackage{xcolor}
\usepackage{scrextend}
\usepackage{multirow}
\usepackage{makecell}
\usepackage{relsize}

\definecolor{changecolour}{rgb}{0, 0.0, 0}

\definecolor{redcolor}{rgb}{.7,0.,0.}

\definecolor{greencolour}{rgb}{0.0, 0.7, 0.0}

\newcommand{\xin}{{{\bf x}^{\text{input}}}}

\newcommand{\sinput}{{s^{\text{input}}}}

\begin{document}

\preprint{APS/123-QED}

\title{Non-parametric power-law surrogates}

\author{Jack Murdoch Moore}
\email{jackmoore@tongji.edu.cn}
\affiliation{%
School of Physical Science and Engineering, Tongji University, Shanghai 200092, People's Republic of China}%
\author{Gang Yan}%
\email{gyan@tongji.edu.cn (corresponding author)}
\affiliation{%
School of Physical Science and Engineering, Tongji University, Shanghai 200092, People's Republic of China}
\affiliation{%
Frontiers Science Center for Intelligent Autonomous Systems, Tongji University, Shanghai, 200092, P. R. China}
\affiliation{%
Center for Excellence in Brain Science and Intelligence Technology, Chinese Academy of Sciences, Shanghai, 200031, P. R. China
}
\author{Eduardo G. Altmann}%
\email{eduardo.altmann@sydney.edu.au}%
 \affiliation{%
 School of Mathematics and Statistics, University of Sydney, New South Wales 2006, Australia
 }%

\date{\today}

\begin{abstract}

Power-law distributions are essential in computational and statistical investigations of extreme events and complex systems. The usual technique to generate power-law distributed data is to 
first infer the scale exponent $\alpha$ using the observed data of interest and then sample from the associated distribution. 
This approach has important limitations because it relies on a fixed $\alpha$ (e.g., it has limited applicability in testing the {\it family} of power-law distributions) and on the hypothesis of independent observations (e.g., it ignores temporal correlations and other constraints typically present in complex systems data). 
Here we propose a constrained surrogate method that overcomes these limitations by choosing uniformly at random from a set of sequences exactly as likely to be observed under a discrete power-law as the original sequence (i.e., regardless of $\alpha$) and by showing how additional constraints can be imposed in the sequence (e.g., the Markov transition probability between states). 
This non-parametric approach involves redistributing observed prime factors to randomize values in accordance with a power-law model but without restricting ourselves to independent observations or to a particular $\alpha$. 
We test our results in simulated and real data, ranging from the intensity of earthquakes to the number of fatalities in disasters.
\end{abstract}

\maketitle


\section{Introduction}

Fat-tailed distributions are one of the most pronounced characteristics from complex systems~\cite{newman2005power,mitzenmacher2004brief,zipf1936psycho}, appearing in paradigmatic models in Statistical Physics (e.g., critical phenomena, preferential attachment processes, self-organized criticality)~\cite{sethna2021statistical,wilson1974renormalization,willis1922some,price1965networks,price1976general,olami1992self,barabasi1999emergence,bak1987self} and in the analysis of a variety of datasets (e.g., city sizes, word-frequencies, earthquake waiting times, and magnitude of disasters)~\cite{simon1955class,zipf1949human,gutenberg1944frequency,bak2002unified}. A key computational tool to investigate all these systems is the generation of synthetic datasets (surrogates) that account for the fat-tail characteristic of the observation or system of interest. 
These surrogate sequences ${\bf x} = x_1, \ldots , x_N$ represent null models which allow tests of properties of the data -- e.g., whether the data is compatible with a specific distribution --  and to make estimations -- e.g., the magnitude of extreme events.

\begin{figure}[htbp]
\centering
\includegraphics[width=0.45\textwidth]{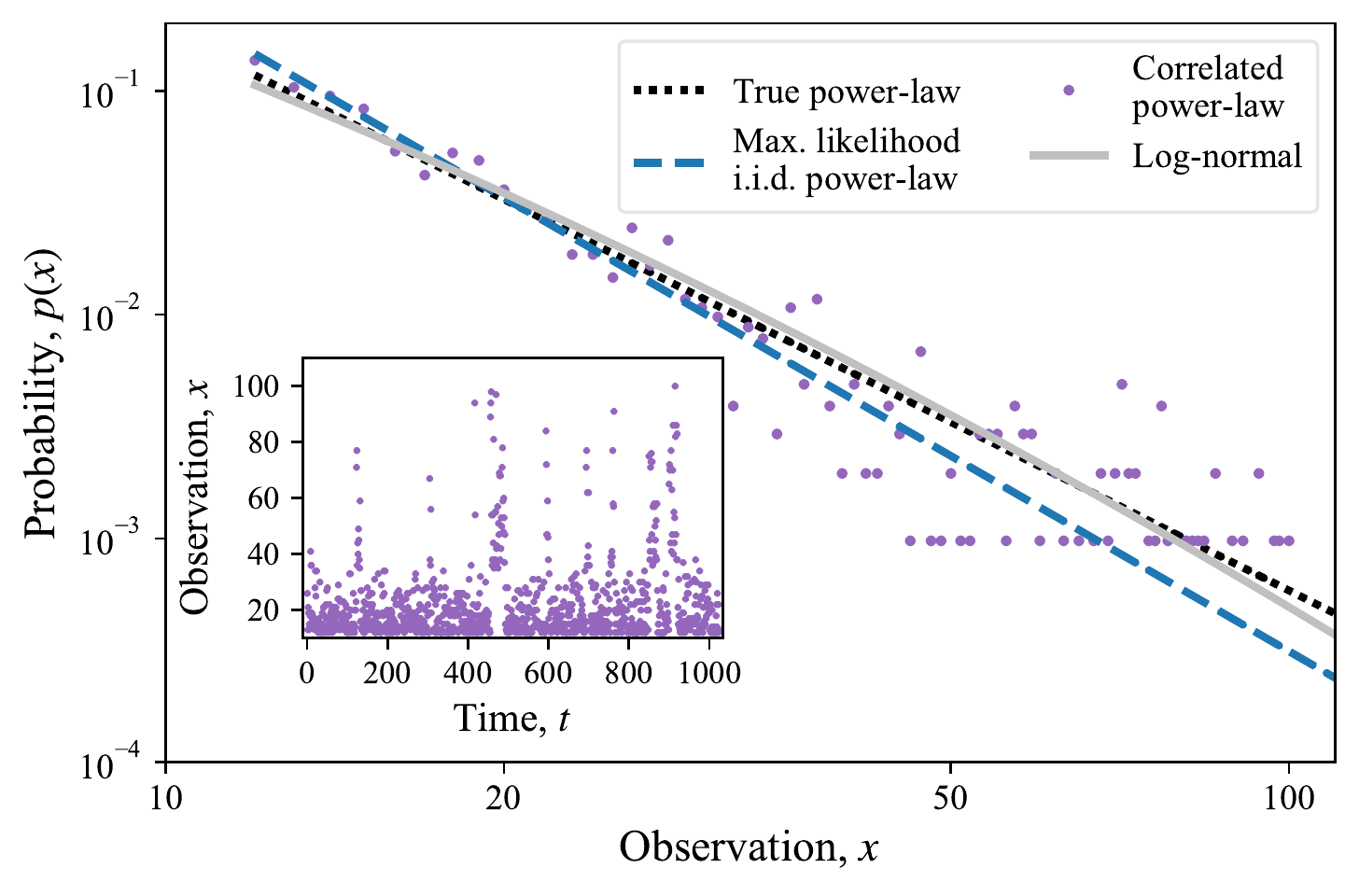}

\caption{
Correlations and limited samples obscure power-laws. A synthetic time series of length $N = 1024$ (see inset) is generated from a correlated (Markov order one) power-law with exponent $\alpha = 2.5$ and lower cut-off $x_{\min} = 12$ (black dotted line, see Appendix \ref{sapp:corr} for details). The time series' empirical distribution is shown (purple filled circles), together with the maximum likelihood fit to an i.i.d. power-law (blue dashed line), and a log-normal distribution (gray solid line, location parameter $\mu = 0$ and scale parameter $\sigma = 1.5$, as defined in Tab. \ref{tab:distributions}).
}
\label{fig:fat_tails}
\end{figure}

The {\it typical approach} to generate surrogates of fat-tailed data is to first fit~\cite{goldstein2004problems,bauke2007parameter,hanel2017fitting} a power-law distribution 
\begin{equation}\label{eq.powerlaw}
p(x) = C x^{-\alpha}
\end{equation}
to the data (or tail, $x \geq x_{\min})$ and then use the maximum-likelihood estimated parameter $\hat{\alpha}$ 
to generate the synthetic dataset~\cite{clauset2009power,deluca2013fitting,corral2019power}. The problems with this approach, which we overcome in this manuscript, are: 
\begin{itemize}
    \item the synthetic dataset arises from a model for the fixed parameter $\hat{\alpha}$ and thus does not represent a null model for power-law distributions in general (arbitrary $\alpha$).
   \item correlations in the data are not accounted for (complex systems data are not i.i.d.)~\cite{bramwell2000universal,corral2008scaling,gerlach2019testing}.
\end{itemize}
Figure~\ref{fig:fat_tails} illustrates how these issues affect data analysis: for small or correlated samples there are substantial differences between the fitted model and the underlying process (estimating scale exponents is hard) and it is difficult to distinguish between different distributions. 
Therefore the two issues above affect directly the perennial debates about the ubiquity of power-law distributions: whether log-normals or power-laws better describe city-size distribution~\cite{eeckhout2004gibrat,levy2009gibrat,eeckhout2009gibrat}, and whether power-law degree distributions are ubiquituous in complex networks~\cite{broido2019scale,voitalov2019scale,zhou2020power-law,falkenberg2020identifying,serafino2021true} and more generally~\cite{stumpf2012critical,holme2019rare,cirillo2020tail,corral2021tail}. Progress on these fundamental debates, that lie at the foundations of complex systems research, requires methods that go beyond the typical approach.  

Surrogates are numerically generated data sequences (time series) that share particular features of the original data (time series) of interest but that randomize other characteristics in accordance with a particular model (or fulfill a certain null hypothesis)~\cite{bradley2015nonlinear,paluvs2007nonlinearity}.
Constrained surrogates fix particular features strictly, matching the exact value observed in the original data, thus allowing us to condition these out and draw principled conclusions about the remaining properties or features which are not being fixed and which we are interested in testing or analyzing~\cite{lancaster2018surrogate,van1998testing}.
Constrained surrogates can represent an entire family of processes, rather than the particular process with the parameters of best fit, and thus have a range of favourable properties explained and explored below. 
Our contribution in this paper is to develop and apply constrained surrogate techniques to sequences with discrete power-law distributions. We will account for different types of correlations without committing to particular values of the scale exponent $\alpha$ (i.e., non-parametrically) but while allowing variation in the values of extreme events and other statistics of interest. We propose a constrained surrogate which redistributes observed prime factors to randomize values while fixing the likelihood of the sequence for all $\alpha$ and avoiding the challenging problem of estimating $\alpha$. We also formulate variants which, in addition, preserve correlations (up to a given Markov order) or other constraints. We then apply these constrained surrogates to artificial and real data, showing that they lead to exact hypothesis tests, provide unbiased and lower-variance estimates of statistics, are more robust to model misspecification, and can accommodate and thus facilitate inference about correlations present in the time series. Finally, we show how the use of our constrained surrogates impacts the conclusions and estimates obtained from different fat-tailed data. Our codes are available in Ref.~\cite{github}.

\section{Surrogate method}

In this section we 
introduce our constrained power-law surrogate methods.
Let ${\bf x} = x_1, \ldots , x_N$ be an ordered sequence of integers, each of which is no less than a lower cut-off $x_{\min}$. Given one such {\em input} sequence $\xin$, we are interested in generating an ensemble $\{{\bf x}_n\}$ of other surrogate sequences ${\bf x}$ constrained to the hypothesis of power-law distribution~Eq.~(\ref{eq.powerlaw}). A simple surrogate is the shuffle surrogate~\cite{theiler1996constrained}, which corresponds to a permutation, chosen uniformly at random, of the input sequence. Another is bootstrapping: sampling uniformly at random from the input sequence with replacement. Other important surrogates {are designed to apply the null hypothesis that observations are a static transformation of a linear process. Surrogates for this hypothesis include the statically transformed autoregressive process~\cite{kugiumtzis2002statically},} amplitude adjusted Fourier transform (AAFT)~\cite{theiler1992testing} {and} iterated AAFT~\cite{schreiber1996improved}, {the last two of which we compare to our proposed surrogate methods in the supplemental material (SM)~\cite{supplemental}}. While these surrogates succeed in capturing features of the fat-tailed distribution in $\xin$, a strong limitation is that they only offer values which have already been observed and therefore they cannot be used to explore unobserved cases (SM~\cite{supplemental}, {Fig.~S1}), such as the extreme events which are particularly important in fat-tailed processes. Constrained surrogates overcome this limitation, while showing other interesting properties not present in the usual approach based on maximum likelihood fitting.

\subsection{Constrained surrogates}\label{ssec:constrained}

Constrained surrogates incorporate hypotheses or constraints by fixing a set of properties $\left\{K\right\}$ to match those observed in $\xin$ such that
\begin{equation}\label{eq.gendef3}
K \left({\bf x}\right) = K\left(\xin\right)
\end{equation}
for all ${\bf x} \in \{ {\bf x}_n\}$.
A constrained property $K$ could be, for example, the number of elements $N$ of the sequence ${\bf x}$, the parity (odd or even) of the sum $\sum\limits_t x_t$, or the truth value of the statement ``$\max\limits_t x_t \leq x_{\max}$'', where $x_{\max}$ is known and fixed. 
Here we are particularly interested in the hypothesis that $\xin$ is generated from a probability distribution with parameter(s) $\alpha$ and the corresponding property $K$ to be constrained is the likelihood function $\mathcal{L}_{{\bf x}}$ 
which maps $\alpha$ to the probability $\mathcal{L}_{{\bf x}}\left( \alpha \right)$ of generating a sequence ${\bf x}$. The constraint Eq.~(\ref{eq.gendef3}) is then the condition
 \begin{equation}\label{eq.gendef4}
 \mathcal{L}_{{\bf x}}\left(\alpha\right)=\mathcal{L}_{\xin}\left(\alpha\right),
 \end{equation}%
 for all ${\bf x} \in \{ {\bf x}_n\}$ and all $\alpha$.
 Constrained surrogates $\{ {\bf x}_n\}$ are obtained by sampling uniformly at random from a collection of (ideally all) sequences $\mathcal{U}\left(\xin\right)$ such that:
\begin{itemize}
    \item[C1.] $\xin \in \mathcal{U} \left( \xin \right)$;
    \item[C2.] $\forall {\bf x} \in \mathcal{U} \left( \xin \right), \mathcal{U} \left( {\bf x} \right) = \mathcal{U} \left( \xin \right)$; and
    \item[C3.] 
    $\forall {\bf x} \in \mathcal{U} \left( \xin \right)$ and for each constrained property $K$, Eq.~(\ref{eq.gendef3}) is satisfied.
\end{itemize}
This procedure factors out the influence of constraints and (unknown) model parameters by conditioning on the output of the map $\xin \mapsto \mathcal{U} \left( \xin \right)$. If we consider (i) $\xin$ as a realization of a discrete random variable ${\bf X}$ drawn from a distribution with parameters $\alpha$ and (ii) the likelihood under the distribution to be one of the constrained properties $K$, then conditions C1-C3 above guarantee that the conditional probability
of ${\bf X}$ given $\mathcal{U}\left( {\bf X} \right)$ is independent of $\alpha$ -- i.e., $P_\alpha\left( {\bf X} \mid \mathcal{U}\left( {\bf X} \right) \right) \equiv P\left( {\bf X} \mid \mathcal{U}\left( {\bf X} \right) \right)$ -- and uniform on $\mathcal{U}\left( {\bf X} \right)$, in agreement with our prescription for constrained surrogates (for a proof, see SM~\cite{supplemental}, {Sec.~I}). {That is, as long as the likelihood under the null hypothesis is one of the fixed quantities, producing a constrained surrogate is equivalent to generating data under the hypothesised process subject to a condition -- the particular value of $\mathcal{U}\left( {\bf X} \right)$ -- which has been observed for the input sequence. In particular, by preserving the likelihood function we ensure that (maximum likelihood) parameter estimations are the same for the input and surrogate sequences. } Constrained surrogates are closely related to the idea of conditioning on a sufficient statistic~\cite{fisher1922mathematical,reid1995roles,van1998testing,lehmann2005testing}.

Favourable properties of constrained surrogates include: 
\begin{itemize}
\item [(i)] they preserve the probability distribution of sequences and, as a consequence,  provide unbiased estimates of the expectation of \emph{any} statistic;

\item[(ii)] the expectation of any sample statistic estimated using the mean over many constrained surrogate datasets realised independently from the same observation has variance no larger than the variance of the original statistic and, as long as this variance is finite and the sample statistic is non-constant over the collection $\mathcal{U}\left({\xin}\right)$ for some $\xin$, strictly smaller.

\item[(iii)] constrained surrogates provide exact~\cite{theiler1997using,besag2013exact,pethel2014exact} (i.e., theoretically supported) hypothesis tests regardless of discriminating statistic or sample length~\cite{theiler1996constrained}. In particular, this allows for hypothesis testing using composite\footnote{A hypothesis is called simple when it is consistent with precisely one process~\cite{theiler1996constrained,small1998correlation,small2005applied,paluvs2007nonlinearity}; more general hypotheses (e.g., hypotheses which do not specify parameter values) are called composite~\cite{theiler1996constrained,small1998correlation,van1998testing,schreiber1999interdisciplinary,schreiber2000surrogate,small2005applied,paluvs2007nonlinearity}} hypotheses and non-pivotal\footnote{A statistic is called pivotal when its distribution is the same for all processes consistent with the tested hypothesis; other statistics are non-pivotal~\cite{theiler1996constrained,small1998correlation,small2005applied,paluvs2007nonlinearity}} test statistics, a requirement for testing power-law distributions (for all $\alpha$) using test statistics of interest.
\end{itemize}
Property (ii) follows from a classical statistical result that gives the total variance of any sample statistic as a positively weighted sum of the variance within and between the level sets of a statistic~\cite{fisher1938statistical}, and which in our case implies, for constrained surrogates and any statistic $s$,
\begin{equation}\label{eq.var}
\mathbb{V} \left[ s\right] = \mathbb{V} \left[\mathbb{E}_{\textrm{surr.}} \left[ s\right] \right] + \mathbb{E} \left[\mathbb{V}_{\textrm{surr.}} \left[ s\right] \right],
\end{equation}
where $\mathbb{E}$ ($\mathbb{V}$) denotes the expectation (variance) over sequences from the original generative process and $\mathbb{E}_{\textrm{surr.}}$ ($\mathbb{V}_{\textrm{surr.}}$) denotes the expectation (variance) over surrogate sequences generated from a single input sequence $\xin$.

\subsection{Constrained power-law surrogates}

\begin{figure}[htbp]
\includegraphics[width=0.48\textwidth]{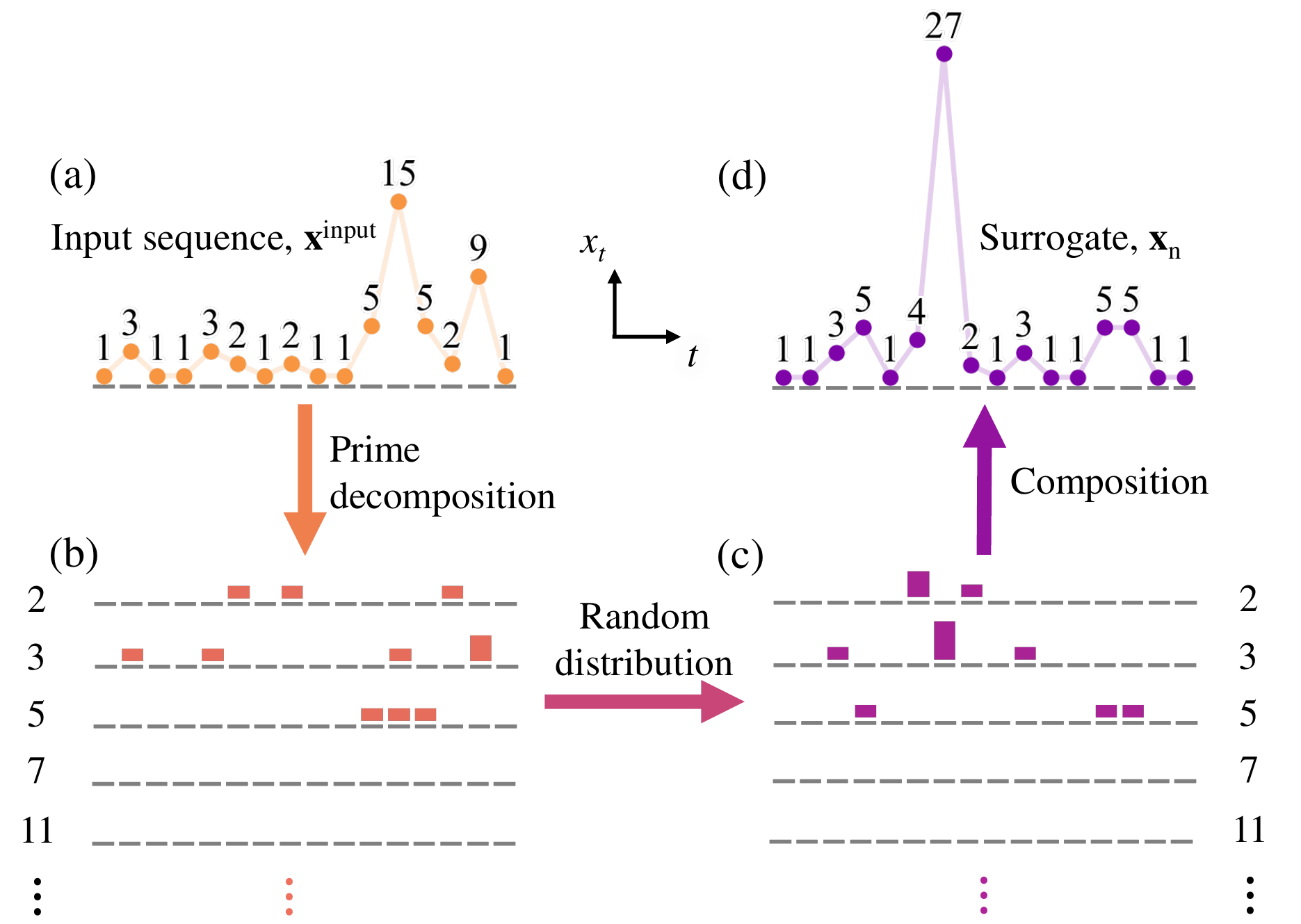}
\caption{Schematic illustration of the process to generate power-law surrogates. (a) to (b): each element $x_t$ of the input sequence $\xin = x_1, \ldots, x_N$ is separately decomposed into its prime factors. (b) to (c): all observed instances of each specific prime factor are randomly redistributed in $t$ {such that each distinct distribution is equally likely}. (c) to (d): the redistributed prime factors are composed into a new surrogate sequence ${\bf x}$. 
}
\label{fig:schematic-iid}
\end{figure}

The constrained power-law surrogate methods we propose here correspond to the null hypothesis that $\xin$ follows a power-law distribution~(\ref{eq.powerlaw}) with (an unknown) exponent $\alpha$\footnote{In contrast, in the typical approach $\alpha$ is fixed to be equal to the maximum-likelihood-estimation of $\alpha$ in $\xin$.}. 
We are interested in constrained surrogates such that for any surrogate sequence ${\bf x} \in \{{\bf x}_n\}$ and for all $\alpha$, the likelihood is the same, i.e., Eq.~(\ref{eq.gendef4}) is satisfied.
For an i.i.d. power-law governed by Eq. (\ref{eq.powerlaw}), the likelihood is
\begin{equation}\notag\label{eq.likelihood}
    \mathcal{L}_{{\bf x}} (\alpha) =  C^{N} \left( \prod\limits_{t=1}^N x_t \right)^{-\alpha}  = C^{N} \left(\prod\limits_{q} q^{n_q}\right)^{-\alpha},
\end{equation}
where $n_q$ is the number of instances of the prime factor $q$ in the product $\prod\limits_{t=1}^N x_t$. It follows that maintaining the likelihood is equivalent to preserving the count of each prime factor which appears in the sequence. By the uniqueness of prime decompositions, when $x_{\min} = 1$, we can choose uniformly at random from among all sequences with the same likelihood by randomly assigning instances of prime factors to elements of the sequence such that, for each distinct prime factor, each possible sequence of counts is equally likely. This procedure satisfies conditions C1-C3 and so produces a constrained surrogate (Sec.~\ref{ssec:constrained}). We illustrate the process in Fig.~\ref{fig:schematic-iid}.

When $x_{\min} > 1$ it is necessary instead to choose uniformly at random from a set of sequences with the same product and each element of which is greater than or equal to the lower cut-off $x_{\min}$. This can be accomplished by: (1) associating with each element of a dataset sufficiently many instances of prime factors that the element cannot dip below $x_{\min}$, (2) randomly allocating instances of prime factors among all elements not associated with any smaller prime factors, and (3) randomising the order of the resulting dataset (see Appendix \ref{sapp:details_alg} for details of {choosing uniformly at random from among all distinct distributions of prime factors by choosing a random weak integer compositions into a fixed number of parts~\cite{nijenhuis2014combinatorial},} generating constrained i.i.d. power-law surrogates with $x_{\min} > 1${,} and fitting the lower cut-off).  

\subsection{Beyond i.i.d.} \label{ssec:markov}
\begin{figure}[htbp]
\includegraphics[width=0.48\textwidth]{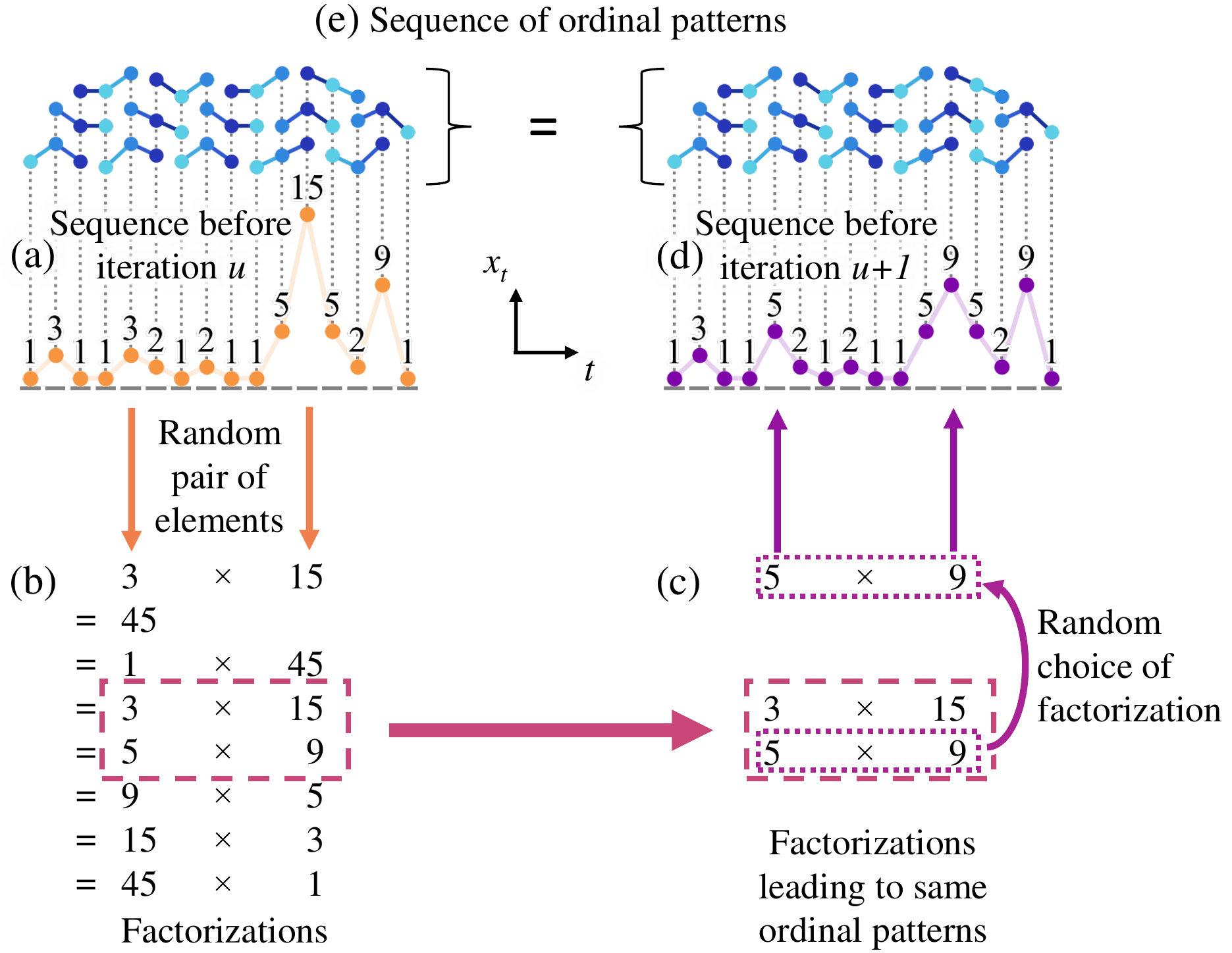}
\caption{Schematic illustration of the process to generate power-law surrogates with additional constraints (ordinal patterns of length $m+1=3$). 
In each iteration $u$ of the algorithm we: (a) choose randomly a pair of elements $x_i,x_j$ ($x_5 = 3,x_{12} = 15$) from the sequence $x_1, \ldots, x_N$; (b) compute all factorizations $x_i'\times  x_j' = x_i \times x_j$ of the product $x_i \times x_j$ ($x_5 \times x_{12} = 45$); (c) choose a single factorization $x_i' \times x_j'$ ($x_5'\times x_{12}' = 5 \times 9$) uniformly at random from all factorizations $x_i'\times x_j'$ which preserve all ordinal patterns in the original sequence ($x_5'\times x_{12}' = 3\times 15, x_5'\times x_{12}' = 5\times 9$); (d) update the sequence by setting $x_i = x_i'$ ($x_5 = x_5' = 5$) and $x_j = x_j'$ ($x_{12} = x_{12}' = 9$). The sequence (a-d) is repeated many times, with each iteration preserving (e) the same sequence of ordinal patterns. }
\label{fig:schematic-op}
\end{figure}

 Here we show how to construct power-law surrogates which go beyond the i.i.d. hypothesis mentioned above and that consider temporal correlations of length $m \geq 1$. We build collections of surrogates $\{{\bf x}_n\}$ that, in addition to the power-law constraint in Eq.~(\ref{eq.gendef4}), are also constrained to have one of the following properties observed in $\xin$:
\begin{enumerate}
    \item The same rank order of all size $m + 1$ subsequence $x_t, x_{t + 1}, \ldots, x_{t + m}$ (length $m + 1$ ordinal patterns%
    \footnote{Given a real time series $x_1, x_2, \ldots, x_N \in \mathbb{R}$, the corresponding sequence of ordinal patterns of length $m + 1$ is the sequence of real $(m + 1)$-vectors $\bm{y}_t, {\bm{y}_{t+1}}, \ldots, \bm{y}_{N - m}$, in which $\bm{y}_t \in \mathbb{R}^{m + 1}$ comprises the rank order of the subsequence $x_t, x_{t + 1}, \ldots, x_{t + m}$, with tied values replaced by their mean rank. Ordinal patterns are discussed in, e.g., Ref.~\cite{bandt2002permutation,amigo2007true,amigo2010permutation}, and Ref.~\cite{mccullough2017regenerating,hirata2019surrogate} discuss (non-power-law) ordinal pattern surrogates (in these references, the definition of ordinal patterns involves a different treatment of tied values).%
    }%
    ).
    \item The same empirical transition probabilities $p(z_{t}|z_{t-1}, \ldots, z_{t-m})$ between states%
    \footnote{The set of all Markov states is a partition $\mathcal{A}$ of the integers greater than or equal to the lower cut-off $x_{\min}$. The Markov state $z_t$ at time $t$ is the unique element of the partition $\mathcal{A}$ which contains the value $x_t$ at time $t$; $x_t \in z_t \in \mathcal{A}$.} 
    constructed as non-overlapping sets of integers (order $m$ Markov).
\end{enumerate}
\color{black}
We produce ordinal pattern power-law surrogates (Case 1) using a Metropolis algorithm which, in the limit of a large number of transitions, samples uniformly at random from a set of sequences $\mathcal{U}\left(\xin\right) \ni \xin$, each of which satisfies Eq.~(\ref{eq.gendef4}) and also exhibits the same sequences of ordinal patterns as the input sequence $\xin$. Because this prescription satisfies conditions C1-C3 (Sec.~\ref{ssec:constrained}), it provides constrained surrogates. The algorithm begins with ${\bf x}= \xin$ and for each iteration:
\begin{itemize}
    \item a pair of distinct observations $x_i$ and $x_j$ (in which $i \neq j$ but, possibly, $x_i = x_j$) is chosen uniformly at random from the sequence $x_1, \ldots, x_N$. 
    \item the observations $x_i$ and $x_j$ are replaced by $x_i'$ and $x_j'$ respectively, where $x_i' x_j' = x_i x_j$ is a factorization chosen uniformly at random from among all which would lead to a sequence which: 
    \begin{itemize}
        \item[(A)] has no element less than the lower cut-off $x_{\min}$; and 
        \item[(B)] exhibits the same sequence (of {length} $N - m$) of ordinal patterns of length $m + 1$. 
    \end{itemize}
\end{itemize}
In Fig.~\ref{fig:schematic-op} we illustrate an iteration of the Metropolis algorithm for generating constrained ordinal pattern power-law surrogates.
The preceding Metropolis algorithm can be adapted to represent alternative assumptions about correlation structure. 
Markov power-law surrogates (Case 2) are also obtained using the Metropolis algorithm above with condition (B) replaced by ``exhibits the same sequence 
of Markov states''. 
Subsequently, the sequence is randomly reordered while preserving empirical transition probabilities between Markov states\footnote{As a result of preserving empirical transition probabilities, the Markov order one surrogate precisely maintains the conditional entropy of order $m$ with respect to the Markov states (see SM~\cite{supplemental}, {Fig.~S2}).}%
~\cite{kandel1996shuffling, van1998testing, pethel2014exact, correa2020constrained}.
In our studies we used $10^5$ transitions of the Metropolis algorithm.

The correlated power-law surrogates defined above can be used to estimate the effective length $m$ of correlations in a power-law sequence. To reach this, we emulate a popular framework for estimating Markov order for non-power-law discrete data~\cite{van1998testing,pethel2014exact,correa2020constrained}. We increase the hypothesised length $m$ of correlations, starting from the i.i.d. case $m = 0$, until we cannot reject with our pre-specified test statistic and level of significance. The lowest value of $m$ for which this occurs provides a quantification of the correlation (or temporal dependencies) present in the sequence. If it appears that no value of $m$ can prevent rejection, then a power-law model together with the type of correlations hypothesised is presumably not an appropriate explanation for the time series.

\color{black}

\section{Applications}

In this section we show how the methods introduced above perform in different applications. This is done by generating surrogates from the following different input sequences $\xin$:
\begin{itemize}
\item i.i.d. power-laws.
\item i.i.d. sequences from three other distributions which can be mistaken for power-laws at small sample lengths: a power-law truncated at its 1/1024 upper quantile, a power-law with exponential cut-off and a discretized lognormal distribution truncated at $x_{\min}$, see Table \ref{tab:distributions}. 
\item Markov sequences with power-law limiting distributions. 
\item real data in the form of records of fatalities from historical epidemics and terrorist attacks, as well as intensities of solar flares, energy release by earthquakes, customers affected by blackouts, and frequencies of words in the novel \emph{Moby Dick} by Herman Melville, see Table \ref{tab:real_data_p}. 
\end{itemize}
We compare constrained power-law surrogates of different Markov order with the conventional method and, where appropriate, with shuffling and bootstrapping on different hypothesis testing and estimation tasks. 

\begin{table*}[!htbp]
	\caption{
		The probability distributions considered. In each case the constant $C$ is defined such that the total probability mass is unity. 
	}
	\label{tab:distributions}%
	\begin{ruledtabular}
		\begin{tabular}{lccc}
			Name & Probability & Support & Parameters\\
			\hline
			\makecell[l]{Power-law
			 }%
			& $p(x) = C x^{-\alpha}$ & $\left\{ x_{\min}, x_{\min} + 1, \ldots \right\}$ & $\alpha = 2.5$, $x_{\min} = 1, 12$\\
			Truncated power-law\footnote{%
            The truncated power-law is a power-law truncated at its 1/1024 upper quantile.} & $p(x) = C x^{-\alpha}$  & $\left\{ x_{\min}, x_{\min} + 1, \ldots, x_{\max} \right\}$ & $\alpha = 2.5$, $x_{\min} = 1$, $x_{\max} = 64$
            \\
			\makecell[l]{
			Power-law with cut-off\\(discretized\footref{footnotedisc})
			}
			& $p(x) = C x^{-\alpha} \exp \left( -\lambda x \right)$ & $\left\{ x_{\min}, x_{\min} + 1, \ldots \right\}$ & $\alpha = 2.5$, $\lambda = 0.01$, $x_{\min} = 1$\\
			\makecell[l]{
			Log-normal\\(Truncated and discretized\footnote{%
		    Initially continuous random variables were rounded to the nearest integer, so that the final probability of observing an integer $x$ is proportional to $\mathlarger{\int\limits_{x - 0.5}^{x + 0.5}} p(y) \mbox{d} y$, where $p$ is the probability density listed above. \label{footnotedisc}}%
			)
			}
			& $p(x) = C \frac{1}{x \sigma \sqrt{2 \pi}} \exp \left[ -\left( \frac{\log x - \mu}{\sqrt{2} \sigma}\right)^2 \right]$ & $\left\{ x_{\min}, x_{\min} + 1, \ldots \right\}$ & $\mu = 0$,
			$\sigma = 1.5$, $x_{\min} = 1,12$
            \\
            Power-law of Markov order $m$\footnote{The generation of correlated power-law sequences is detailed in Appendix \ref{sapp:corr}.\label{footnotecorr}}
			& $p(x) = C x^{-\alpha}$ & $\left\{ x_{\min}, x_{\min} + 1, \ldots \right\}$ & $\alpha = 2.5$, $x_{\min} = 1$
            \\
			\makecell[l]{
			Power-law with correlation\\
			(Lyapunov) time $\tau$\footref{footnotecorr}}
			& $p(x) = C x^{-\alpha}$ & $\left\{ x_{\min}, x_{\min} + 1, \ldots \right\}$ & $\alpha = 2.5$, $x_{\min} = 1$
		\end{tabular}
	\end{ruledtabular}
\end{table*}

\subsection{Hypothesis testing}

In hypothesis testing, a null hypothesis is rejected when the \emph{p-value} -- i.e., the probability (under this hypothesis) of the value of a discriminating statistic (computed from the input sequence $\xin$) -- is smaller than a predetermined threshold (or nominal size parameter, typically set to $0.05$ or $0.1$). Surrogates provide a computationally efficient procedure to perform hypothesis testing because the probability of different discriminating statistics can be estimated by computing their value in the surrogate ensemble $\{{\bf x}_n\}$.
Here we use different discriminating statistics -- the mean (an average), variance (an indicator of spread), maximum (the most extreme event observed), conditional entropies of order one and two (the conditional entropy of order $m + 1$ quantifies Markov properties of order $m + 1$ and is appropriate for assessing the null hypothesis that data are Markov of order $m$; see Appendix \ref{sapp:con_ent}), and the KS-distance, relative to its maximum likelihood parameter $\hat{\alpha}$, of the part of the dataset {no less than} the lower cut-off (a popular way to assess goodness-of-fit to a power-law~\cite{clauset2009power}) -- and one-sided hypothesis tests 
{(see Appendix \ref{sapp:imp_hyp} for details on the implementation of hypothesis tests)}.
%
To quantify the efficiency of different surrogate methods in each such hypothesis test, we will compute two key quantities:

\begin{itemize}
    \item The \emph{size} of a test is the rate of rejection of the null hypothesis when it is indeed true (incorrect rejection). 
    A test is \emph{exact} when its size equals the predetermined nominal size. 
    \item The \emph{power} of a test as the rate of rejection of the null hypothesis when this hypothesis is incorrect, which depends also on the process underlying the input sequence.
\end{itemize}

\begin{figure}[htbp]

\includegraphics[width=0.45\textwidth]{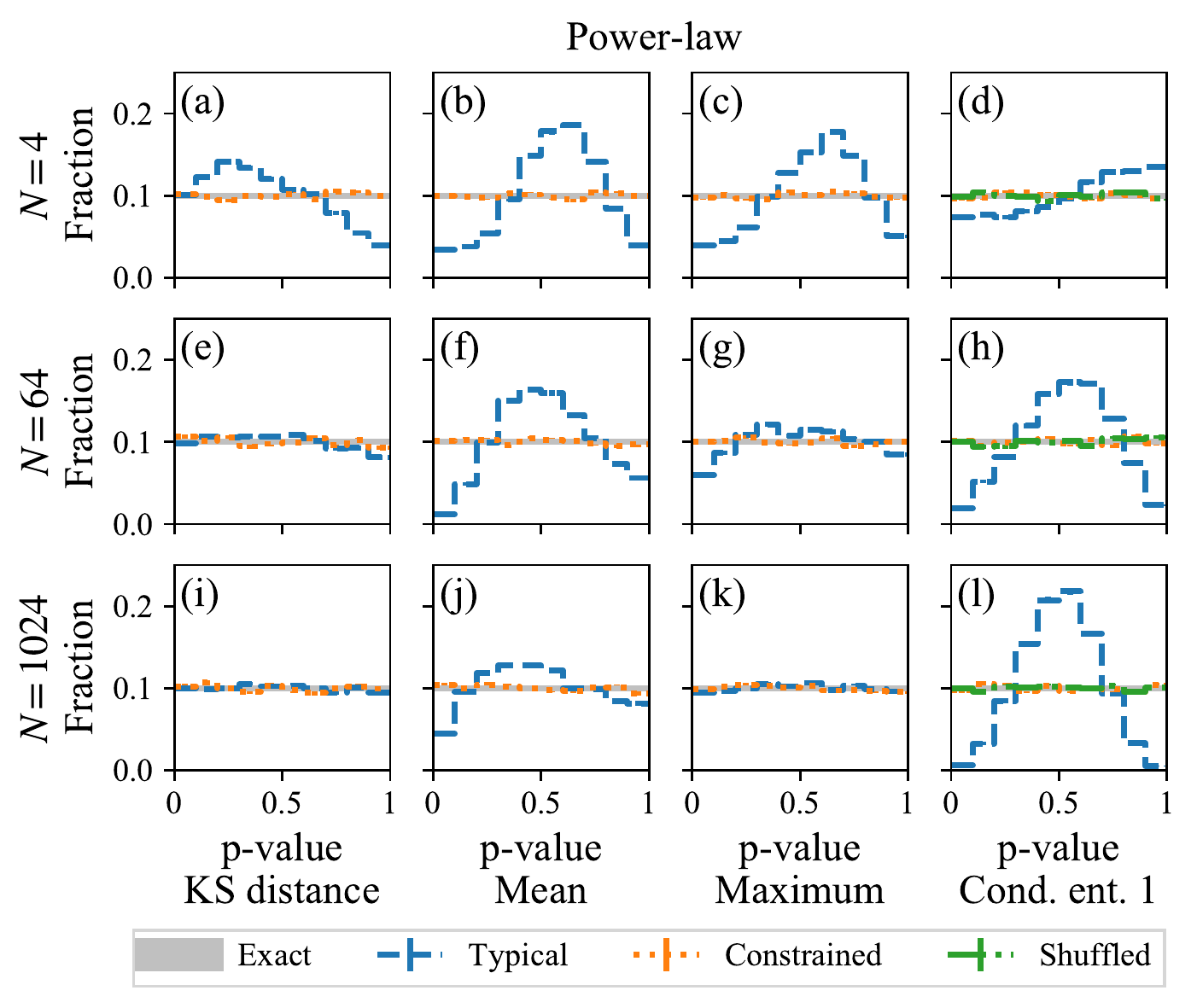}

\caption{Constrained surrogates lead to flat distributions of p-values. Distributions of p-values estimated from $10,000$ hypothesis tests each using an independent sample of length $N$ from an i.i.d. power-law with scale exponent $\alpha = 2.5$ and lower cut-off $x_{\min} = 1$. Each test uses nine typical, constrained or shuffle surrogates, with nominal size 10\%, and uses as test statistic the KS-distance, mean, maximum or conditional entropy of order one (from left to right). The gray line spans one standard error above and below the expected fraction of p-values in each interval, and error bars of other lines correspond to standard error but are smaller than the line width. 
}
\label{fig:p_vals}
\end{figure}

\begin{figure}[htbp]
\includegraphics[width=0.45\textwidth]{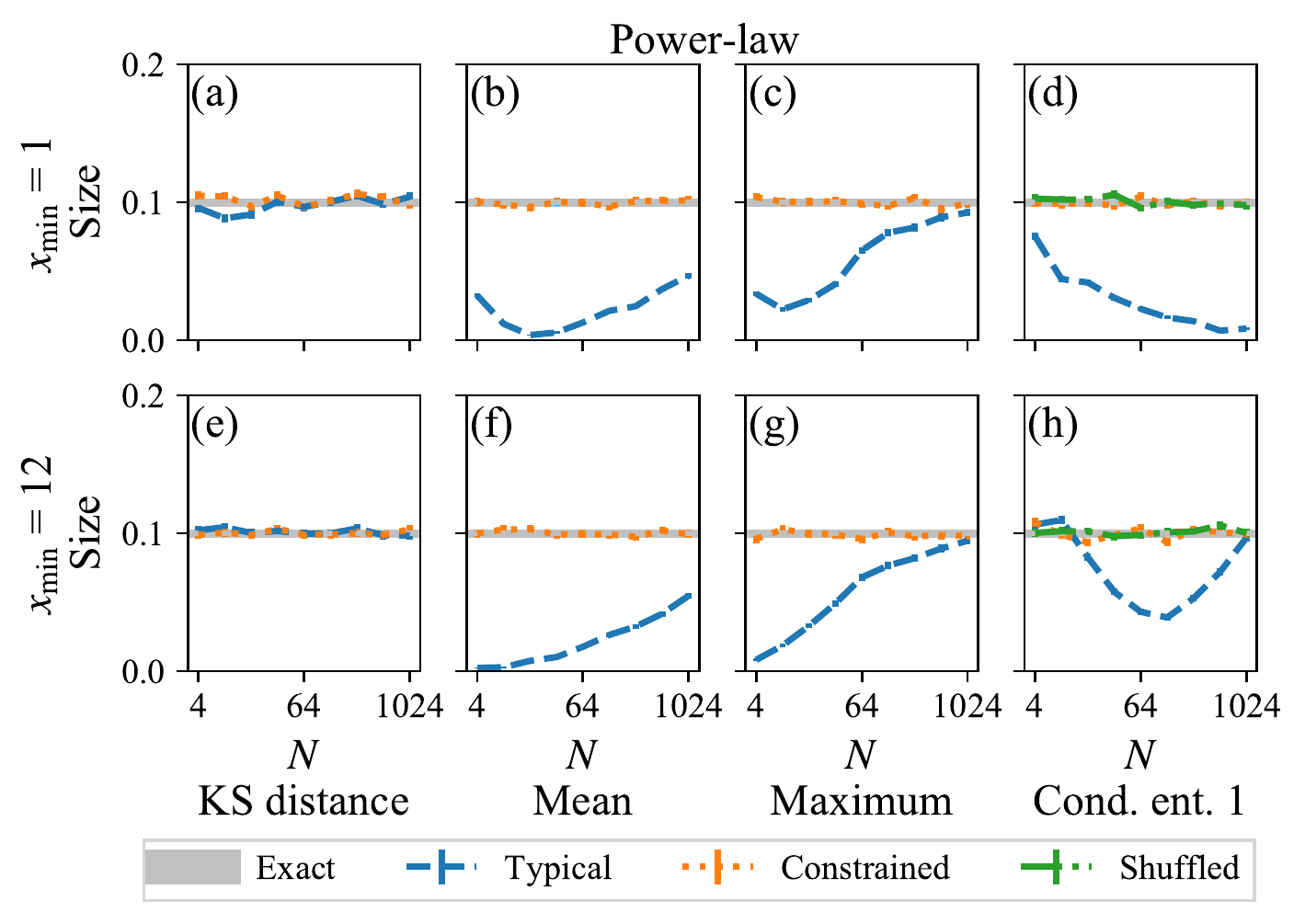}

\caption{Constrained surrogates lead to exact tests (true size equals nominal size), regardless of sample length or discriminating statistic. Sizes estimated from $10,000$ hypothesis tests, each using an independent sample of length $N$ from an i.i.d. power-law with scale exponent $\alpha = 2.5$. Tests use nine typical, constrained or shuffle surrogates, with nominal size 10\%, and use as test statistics the KS-distance, mean, maximum or conditional entropy of order one (from left to right). The gray line spans one standard error above and below the nominal size, and the error bars of other lines correspond to standard error but are at most only slightly larger than the line width.
}
\label{fig:size}
\end{figure}

First we test the i.i.d. power-law hypothesis and, using shuffle surrogates, a more general i.i.d. hypothesis, for input sequences generated from an i.i.d. power-law. In this case, the distribution of p-values is expected to be flat~\cite{hung1997behavior,bhattacharya2002median}. Figure \ref{fig:p_vals} shows this is obtained for constrained power-law surrogates, regardless of discriminating statistic or sample length, but not for typical power-law surrogates. Even when the KS-distance is used as a discriminating statistic, as recommended in Ref.~\cite{clauset2009power}, typical surrogates lead to small but statistically significant deviations from uniformity which are particularly relevant for small sample sizes $N$. Figure~\ref{fig:size} confirms these results by showing how the size of different tests scale with $N$. The constrained power-law surrogates have an exact size for all discriminating statistics, while the typical approach shows pronounced deviations from the desired nominal value. When the conditional entropy of order one is used as a discriminating statistic, the traditional approach shows a deviation from uniformity that even increases with $N$. Differences between true and nominal size arise for typical surrogates because the method allows large variation in the parameter of best fit which, in turn, lead to excessive variability in discriminating statistics~\cite{theiler1996constrained}.

\begin{figure}[htbp]
\includegraphics[width=0.45\textwidth]{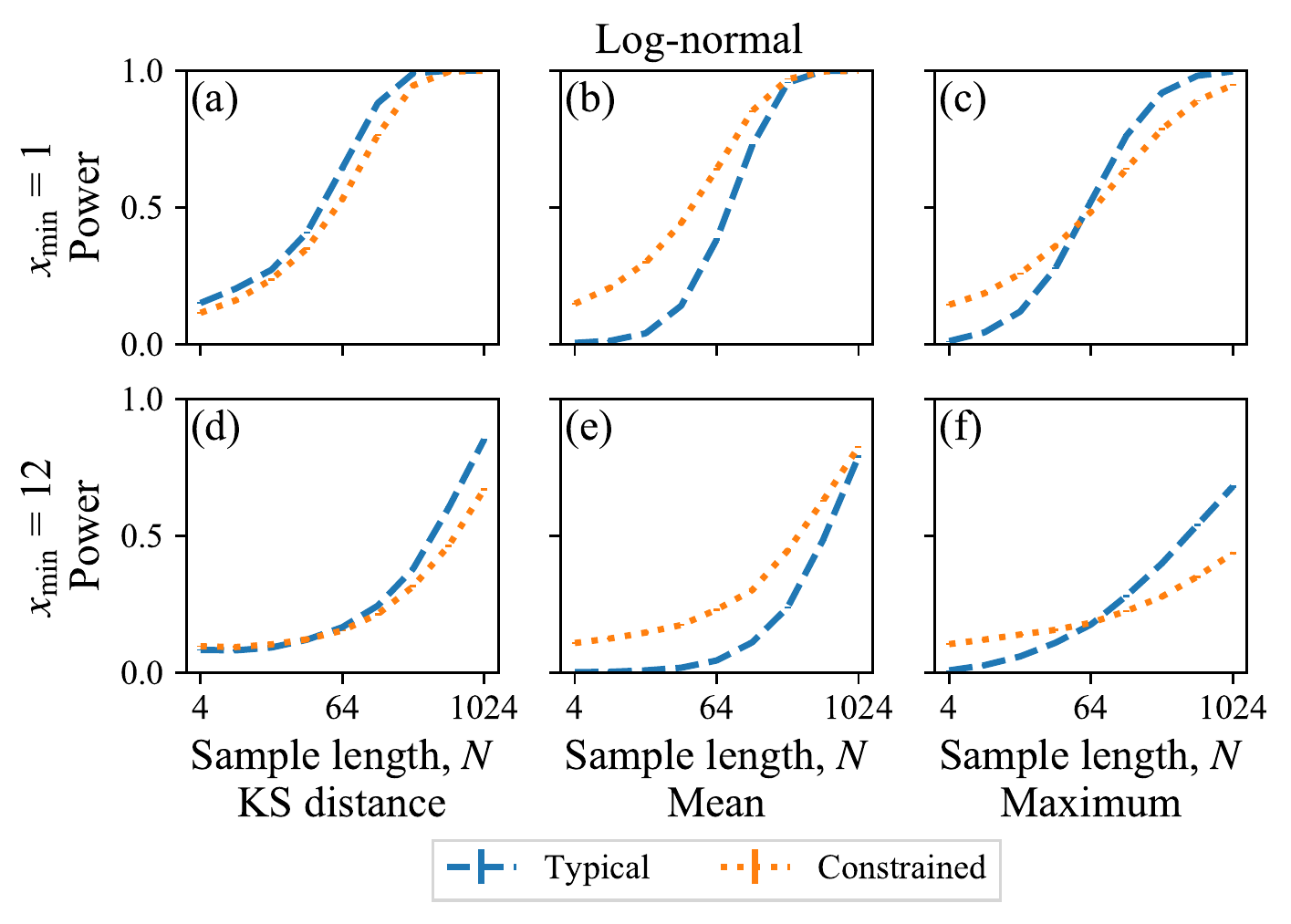}
\caption{Constrained surrogates can increase power for small sample lengths. Powers estimated from $10,000$ hypothesis tests (power-law null hypothesis) each using an independent sample of length $N$ from an i.i.d. log-normal distribution with location parameter $\mu = 0$ and shape parameter $\sigma = 1.5$. Tests use nine typical or constrained surrogates, with nominal size 10\%, and use as discriminating statistics the KS-distance, mean or maximum (from left to right). Error bars show standard error and are at most only slightly larger than the line width.
}
\label{fig:power_log_norm}
\end{figure}

As a next test we consider sequences which actually arose from an i.i.d. log-normal distribution, and compare the rate at which hypothesis tests based on typical and constrained surrogates can rule out the hypothesis of an i.i.d. power-law origin. Figure \ref{fig:power_log_norm} shows that when the KS-distance is used as a statistic, typical surrogates provide slightly greater power, but that for other (non-pivotal) statistics constrained surrogates often lead to similar or higher power, as reported more generally in Ref.~\cite{theiler1996constrained} for nonpivotal statistics.  Most importantly, constrained surrogates seem to perform systematically better in the crucial case of small sample size $N$. In the limit of small sample length $N$, the power of tests based on typical surrogates can approach zero, but the power arising from constrained surrogates approaches the nominal size.

Finally, we apply hypothesis tests to fat-tailed empirical data recording deaths from historical epidemics, the numbers of customers affected by blackouts, word frequencies in the novel \emph{Moby Dick} by Herman Melville, the number of deaths as a direct result of terrorist attacks, intensities of solar flares and energies released by earthquakes. In Tab. \ref{tab:real_data_p} we show that the p-values associated with the hypothesis that real data are i.i.d. power-law above a lower cut-off can vary considerably with statistic and surrogate method. 
The difference between surrogate methods is especially noticeable for the ``Terrorism'' data when the mean or variance is used as a discriminating statistic.

\begin{table*}[!htbp]
	\caption{
		Constrained and typical power-law surrogates can lead to different conclusions about empirical data. The p-values identified via tests using the KS-distance, mean, variance and maximum and $999$ surrogates of the hypothesis that the $N_G$ elements of a dataset which exceed the fitted lower cut-off $\hat{x}_{\min}$ (Appendix \ref{sapp:details_alg}) arose i.i.d. under a power-law~\cite{clauset2009power}. Values which, ignoring issues associated with multiple tests, would allow rejection at the 10\% level of significance~\cite{clauset2009power} are shown in {\bf bold}. The lower cut-off $x_{\min}$ is fitted to the value which minimises the KS-distance between the empirical pdf and the maximum likelihood discrete power-law on $x \geq x_{\min}$~\cite{bauke2007parameter,clauset2009power,broido2019scale}, and surrogates are conditioned on fitted lower cut-off $\hat{x}_{\min}$ (Appendix~\ref{sapp:details_alg}). When continuous datasets are discretized or converted to units of $10^3$ (in the cases indicated in the footnotes to this table), we are considering power-law models for the coarsened sequences rather than the data to their original precision.
	}
	\label{tab:real_data_p}%
	\begin{ruledtabular}
		\begin{tabular}{lrrrcccccccc}
			&&&&\multicolumn{4}{c}{Typical}&\multicolumn{4}{c}{Constrained}\\
			\cline{5-8} \cline{9-12}
			Name & \multicolumn{1}{c}{$N$} &  \multicolumn{1}{c}{$\hat{x}_{\min}$} &  \multicolumn{1}{c}{$N_G$} & $p_{\textrm{KS}}$ & $p_{\textrm{Mean}}$ & $p_{\textrm{Var.}}$ & $p_{\textrm{Max.}}$ & $p_{KS}$ & $p_{\textrm{Mean}}$ & $p_{\textrm{Var.}}$ & $p_{\textrm{Max.}}$ \\
			\hline
            \makecell[l]{Diseases\footnote{``Diseases'' corresponds to average estimates in units of $10^3$ of fatalities due to historical epidemics, rescaled as outlined in Ref.~\cite{cirillo2020tail}.}}
            & 72 & 2317 & 27 & 0.834 & 0.363 & 0.343 & 0.318 & 0.282 & 0.388 & 0.396 & 0.300 \\
            Blackouts\footnote{``Blackouts'' comprises the numbers of customers, in $10^3$ and rounded to the nearest integer, affected by electrical blackouts in the United States between 1984 and 2002~\citep{newman2005power,clauset2009power}:\newline \url{https://aaronclauset.github.io/powerlaws/data/blackouts.txt}}
            & 211 & 235 & 57 & 0.883 & 0.435 & 0.390 & 0.476 & 0.557 & 0.369 & 0.466 & 0.581  \\
            Terrorism\footnote{``Terrorism'' lists the number of deaths as a direct result of terrorist attacks which took place between February 1968 and June 2006~\citep{clauset2007frequency}:\newline \url{https://aaronclauset.github.io/powerlaws/data/terrorism.txt}}
            & 9,101 & 12 & 547 & 0.679 & 0.680 & 0.743 & 0.763 & 0.939 & \textbf{0.027} & \textbf{0.023} & 0.293 \\
            Flares\footnote{``Flares'' lists the peak gamma-ray intensity of solar flares, in counts per second, made from a particular satellite between 1980 and 1989~\cite{newman2005power}:\newline \url{https://aaronclauset.github.io/powerlaws/data/flares.txt}\newline
            Intensities less than 323 counts per second were discarded~\cite{clauset2009power}, then the time series was divided by twice this lower cut-off, to obtain a sequence of real numbers bounded below by 0.5. Finally, each element was rounded to the nearest integer, rounding up when two integers were equally close.}
            & 1,711 & 1 & 1,711 & \textbf{0.008} & \textbf{0.058} & \textbf{0.012} & \textbf{0.003} & \textbf{0.006} &  \textbf{0.065} &\textbf{0.011} & \textbf{0.001} \\
            Words\footnote{``Words'' comprises the count of unique words in the novel \emph{Moby Dick} by Herman Melville~\cite{newman2005power,clauset2009power}:\newline \url{https://aaronclauset.github.io/powerlaws/data/words.txt}}
            & 18,855& 7 & 2,958 & 0.712 & 0.191 & \textbf{0.097} & 0.141 & 0.346 & 0.161 & 0.122 & 0.169 \\
            Earthquakes\footnote{``Earthquakes'' records the approximate energies released by the $59\ 555$ earthquakes of magnitude at least 2.0~\cite{corral2004long} detected in southern California between 1981 and 2000~\cite{yang2012computing}:\newline             \url{https://scedc.caltech.edu/data/alt-2011-yang-hauksson-shearer.html}\newline
            Each earthquake magnitude $M$ was converted to an approximate energy $E$, in Joules, using the formula~\cite{baath1966earthquake}
            \begin{align}
                \log_{10} E = 5.24 + 1.44 M. \notag
            \end{align}
            These energies were divided by twice the energy required for an earthquake of magnitude 2.0, to obtain a sequence of real numbers bounded below by 0.5. Finally, each element of this sequence of energies was rounded to the nearest integer.
            }
            & 59,555 & 1 & 59,555 & \textbf{0.001} & 0.798 & 0.752 & 0.714 & \textbf{0.003} & 0.913 & 0.866 & 0.840
		\end{tabular}
	\end{ruledtabular}
\end{table*}

\subsection{Estimation}

Now we show the advantages of using constrained surrogates to estimate quantities of interest. We are particularly interested in extreme events (e.g., expected sample maxima)
because of their significance in processes with fat-tailed distributions such as power-laws and, throughout this section, employ statistics sensitive to the tail of the distribution. In contrast to shuffling and bootstrapping, constrained surrogates and the typical approach allow for the estimation of the probability of unobserved (extreme) events. The benefit of constrained surrogates over the typical approach is that they avoid biasing estimations of extreme events which, as we will see below, is particularly relevant for small $N$.

\begin{figure}[htbp]
\includegraphics[width=0.45\textwidth]{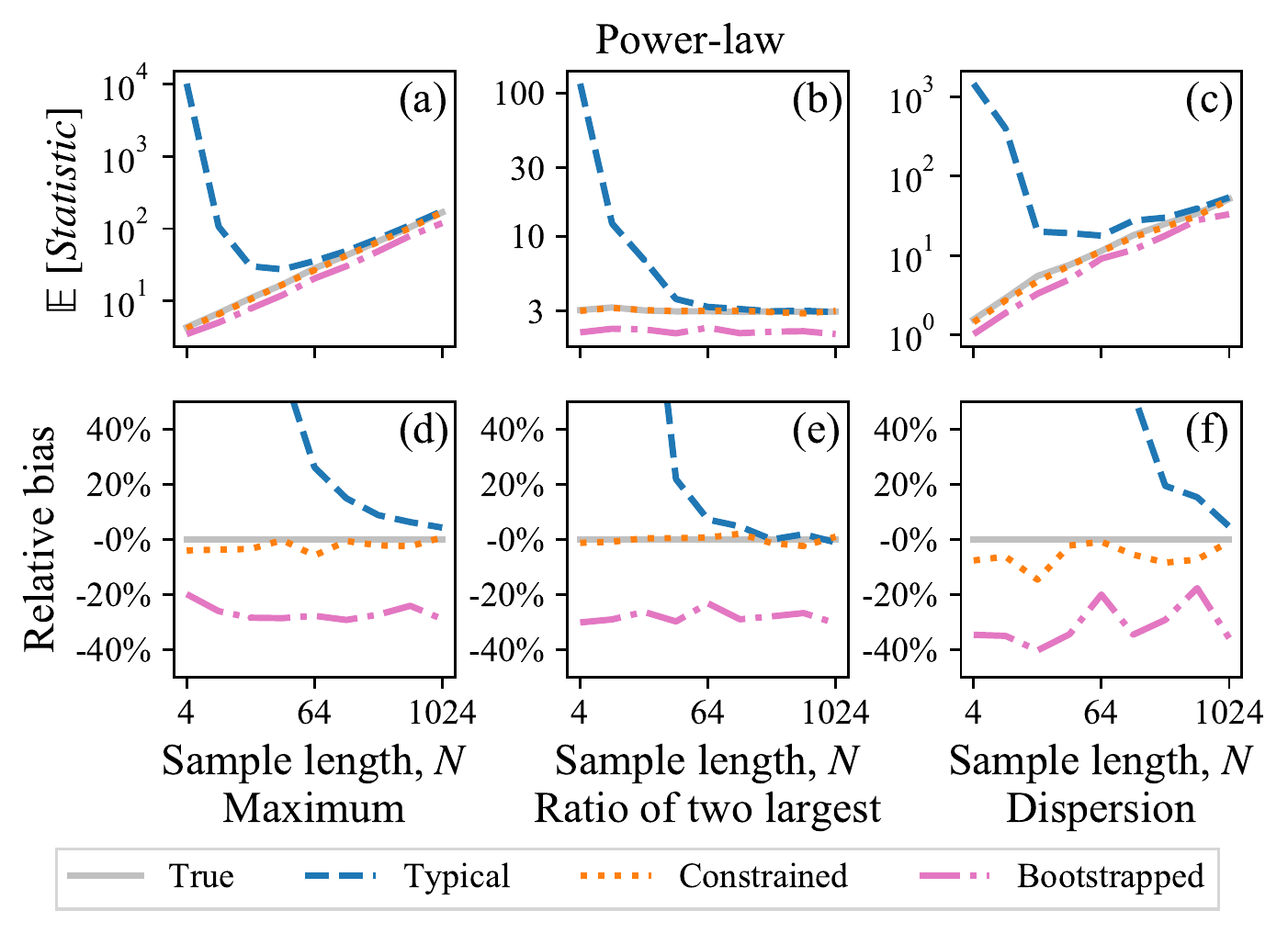}
\caption{
Constrained surrogates do not bias sample statistics. (a-c) Expectation $\mathbb{E}$ of sample statistics and (d-f) relative bias of the calculated expectation. The expectation
$\mathbb{E}$ is calculated using $10,000$ independent input sequences $\xin$, each comprising $N$ values drawn i.i.d. from a power-law distribution with $\alpha = 2.5$, using each input sequence to generate $100$ surrogates. The surrogates considered are either generated independently from the true underlying process (solid gray), typical, constrained or bootstrapped. Surrogates have the same length $N$ as the original sequence (horizontal axis). Results for the sample maximum, ratio of the two largest observations, and index of dispersion are shown in each column and were computed in sequences of length $N$. 
}
\label{fig:mean_var_statistics}
\end{figure}

First, we consider independently generating power-law sequences, and for each surrogate sequence estimating the sample maximum, ratio of the two largest values, and index of dispersion (the ratio of variance to mean~\cite{perry1979power}). For each statistic $s$ we also calculate the relative bias $\left(\tilde{\mathbb{E}}[s] - \mathbb{E}[s]\right)/\mathbb{E}[s]$ in the expectation $\tilde{\mathbb{E}}[s]$ computed using the considered surrogate method.
Figure \ref{fig:mean_var_statistics} shows that constrained surrogates provide an unbiased estimate in all cases, in contrast to bootstrapping and the typical approach. This advantage is particularly important in the relevant case of  small sample size $N$ (e.g., the expectation of sample maximum is much larger when based on typical surrogates, and smaller when based on bootstrapping). Constrained power-law surrogates also reduce {finite} variance, as predicted by Eq.~(\ref{eq.var}), without introducing bias (see SM~\cite{supplemental}, {Fig.~S3}{, where, in addition to statistics investigated in the main paper, we consider a measure of inequality or heterogeneity called the coefficient of variation~\cite{champernowne1998economic}}).

\begin{figure}[htbp]
\includegraphics[width=0.45\textwidth]{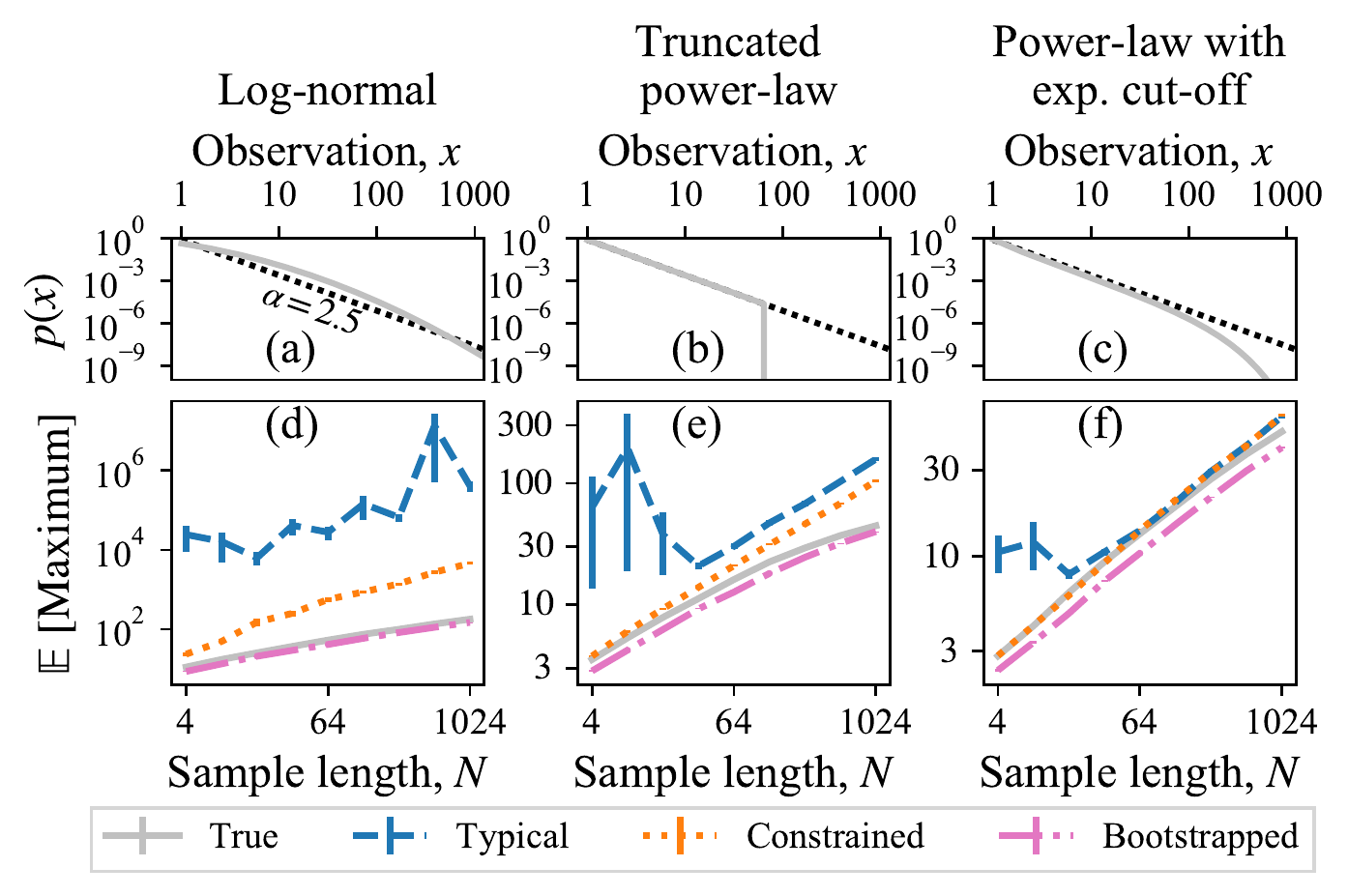}
\caption{
Constrained surrogates ameliorate overestimation of the expected maximum arising from misapplication of a power-law model. The expected maximum in $N$ samples was estimated as the mean over $10,000$ independent realizations of input sequences $\xin$ drawn i.i.d. from three distributions (left to right): log-normal, power-law truncated at its 1/1024 upper quantile, and a power-law with an exponential cut-off. In the upper panels (a-c), the probability mass functions $p(x)$ of these distributions (solid gray line) is compared with that of a power-law with scale exponent $\alpha = 2.5$ and lower cut-off $x_{\min} = 1$ (densely dotted black line). In the lower panels (d-f), the true values of the expected maxima are compared to the ones estimated using three surrogate methods: typical, constrained, and  bootstrap. Error bars show standard error.
}
\label{fig:max_other_data}
\end{figure}

Now we consider time series which are not power-law distributed, but which could reasonably be misidentified as such at small sample length (log-normal distribution, truncated power-law or power-law with exponential cut-off, see Table \ref{tab:distributions}). 
Although the model will not be optimal for these time series, in the absence of knowledge of the true underlying process decision-makers may decide to apply power-law surrogates, especially when trying to allow for the possibility of extreme events. Figure~\ref{fig:max_other_data} shows that constrained surrogates also improve estimates of the expectation of statistics in this context. In particular, the overestimation of the sample maxima is not as extreme as the one obtained using the typical approach. For these non-power-law distributions, bootstrapping, which does not apply a power-law model, provides estimates of maximum closer to the true expected maximum than estimates arising from either typical or constrained power-law surrogates. However, bootstrapping cannot produce new values, with the consequence that resulting maxima are always less than or equal to the maximum of the original observation.

\begin{figure}[htbp]
\includegraphics[width = 0.45\textwidth]{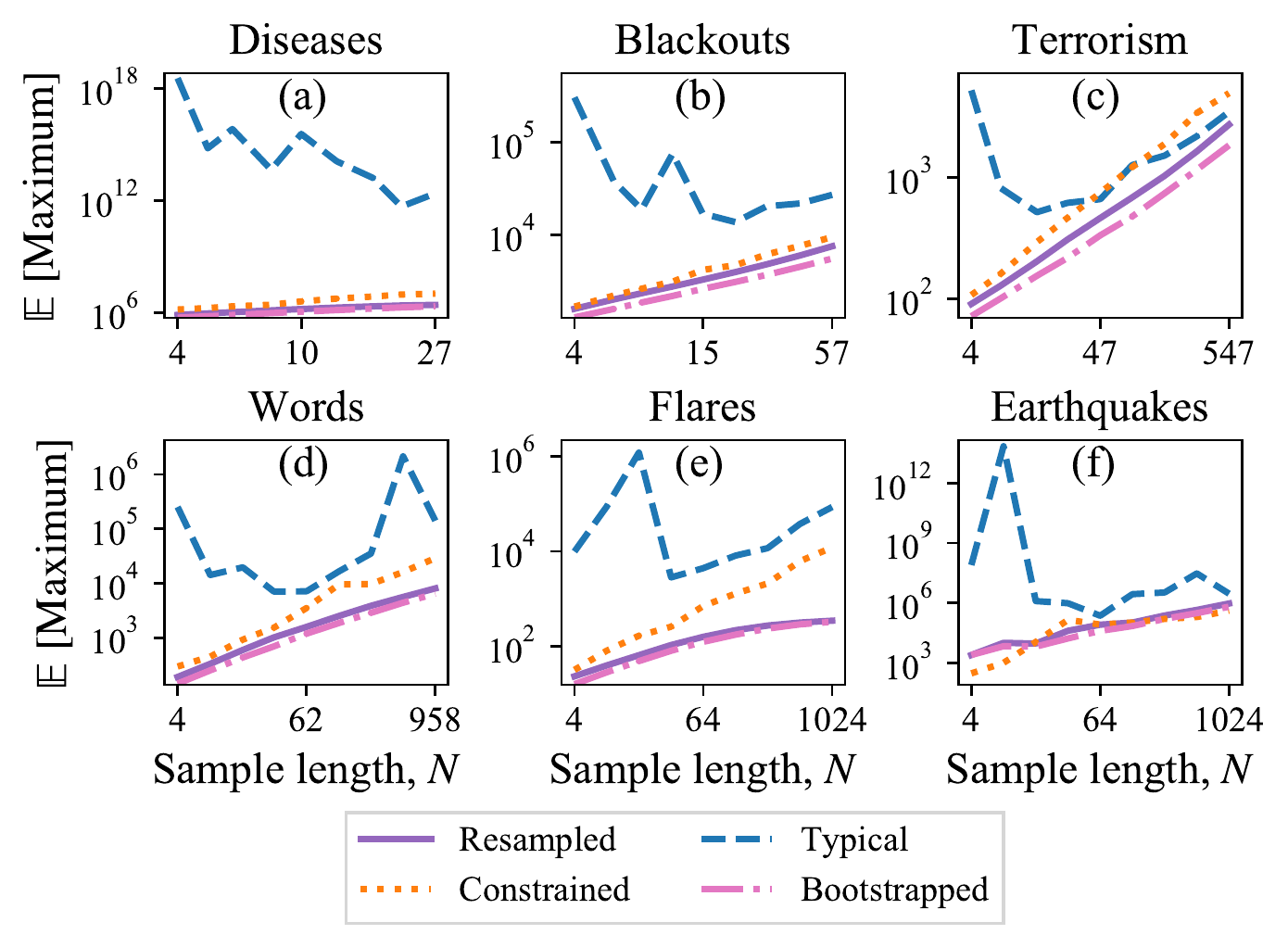}
\caption{
Constrained surrogates avoid extreme values of expected maximum when applying a power-law model to empirical observations (Table \ref{tab:real_data_p}). Expectations of sample maximum estimated as the mean over $10,000$ independent samples of length $N$ drawn i.i.d. from the original data (solid purple) or a typical, constrained or bootstrap surrogate produced from this sample. Only values of the original data which equal or exceed the fitted lower cut-off $\hat{x}_{\min}$ are considered. 
}
\label{fig:max_real_data_sets}
\end{figure}

Finally, we consider again fat-tailed empirical data of varied origin. In Fig.~\ref{fig:max_real_data_sets}, we compare the predictions made by typical and constrained power-law surrogate methods of the expected maximum of random subsamples of varied length $N$. This includes estimates under an i.i.d. power-law model of the expected maximum number of people killed in a single event among the next $N$ epidemics or terrorist attacks. Typical surrogates often produce estimates of expected sample maxima which are alarmingly large. Conversely, bootstrapping, because it cannot provide unobserved values, systematically underestimates the expectation of sample maximum. Constrained surrogates provide a compromise which avoids systematic underestimation but leads to sample maxima which, in expectation, are usually closer to the expected maximum of samples drawn i.i.d. from the original data than are the sample maxima of typical surrogates (though not as close as the systematic underestimates available from bootstrapping).

\subsection{Correlated data}
In this section we explore power-law and empirical datasets that are not i.i.d. We show that correlations impact sample statistics and hypothesis tests based on traditional surrogates, but can be accommodated by constrained surrogates. We focus consistently on the possibility that empirical data are correlated, rather than the alternative explanation of deviations from i.i.d which could be provided by non-stationarity.

\begin{figure}[htbp]
\includegraphics[width=0.45\textwidth]{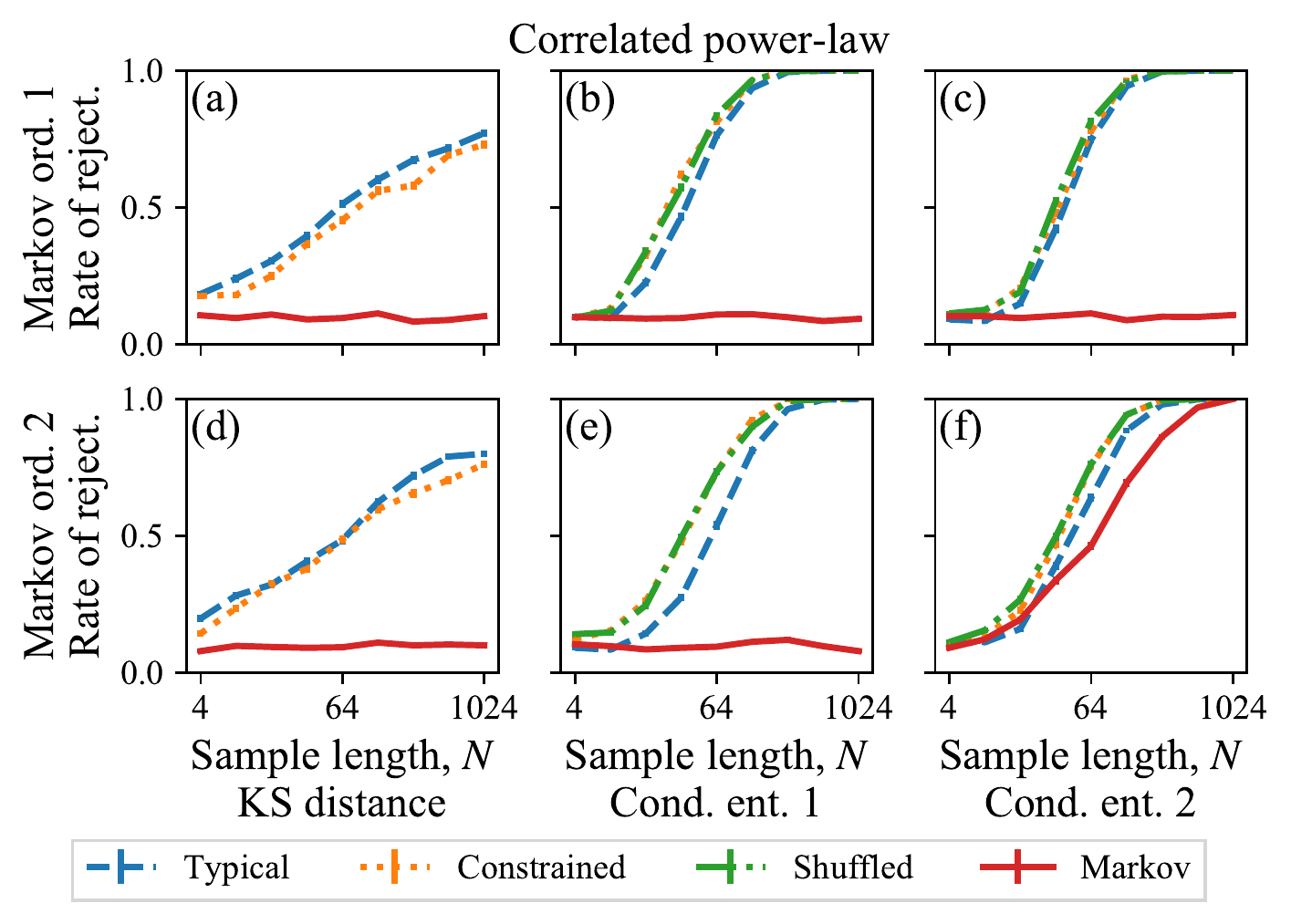}
\caption{
Constrained Markov order power-law surrogates can reduce correlation-based rejection of power-law hypotheses. Rates of rejection estimated from $1,000$ hypothesis tests, each using a sample of length $N$ from a power-law which has scale exponent $\alpha = 2.5$, lower cut-off $x_{\min} = 1$ and Markov order (a-c) one or (d-f) two (Appendix \ref{sapp:corr}). Tests use nine typical, constrained, shuffled or constrained Markov order one power-law surrogates, have a nominal size 10\%, and use as test statistics the KS-distance or conditional entropy of order one or two. Note that the conditional entropy of order $m + 1$ is an appropriate statistic for testing for Markov order $m$ (Appendix \ref{sapp:con_ent}). Markov power-law surrogates use the same Markov states as the original time series. Error bars show standard error and are at most only slightly larger than the line width.
}
\label{fig:power_corr}
\end{figure}

First, in Fig. \ref{fig:power_corr}, we consider data with a stochastic component: these are indeed power-law, but also Markov of order one or two (Appendix \ref{sapp:corr}). Whether the KS-distance or a conditional entropy is used as a discriminating statistic, typical and constrained power-law surrogates - which enforce an i.i.d. power-law hypothesis - both lead to high rates of rejection of the power-law hypothesis (at least, for long samples). This rejection occurs not because the sequences are not power-law, but because they are not i.i.d. The same is true of shuffle surrogates, designed to apply an i.i.d. hypothesis without utilizing a power-law model, when the conditional entropy is used as a discriminating statistic. Constrained Markov order power-law surrogates, designed to enforce the null hypothesis that an observed sequence is power-law and Markov of order one with a given set of Markov states, lead to different behaviour: constrained Markov order power-law surrogates avoid inappropriate (and provide appropriate) rejections. When the original power-law data are Markov of order one, a constrained power-law Markov order one surrogate (with correctly chosen Markov states) leads to a rate of rejection which closely matches the nominal size of the test. When the original power-law sequence is Markov of order two, and the constrained power-law Markov order one surrogate is used, rates of rejection are, once again, close to the nominal size when either the KS-distance or conditional entropy of order one is used as a discriminating statistic. This result is reasonable, because the KS-distance and conditional entropy of order one are not designed to be sensitive to Markov properties of order greater than one. When the conditional entropy of order two, which is sensitive to order two Markov properties, is used as a discriminating statistic, the rate of rejection correctly approaches unity as the sample length increases. Constrained Markov order power-law surrogates maintain similar advantages across a wider range of discriminating statistics and competing surrogate methods than shown here (see SM~\cite{supplemental}, {Fig.~S4}).

\begin{figure}[p]
\centering
\includegraphics[width=0.45\textwidth]{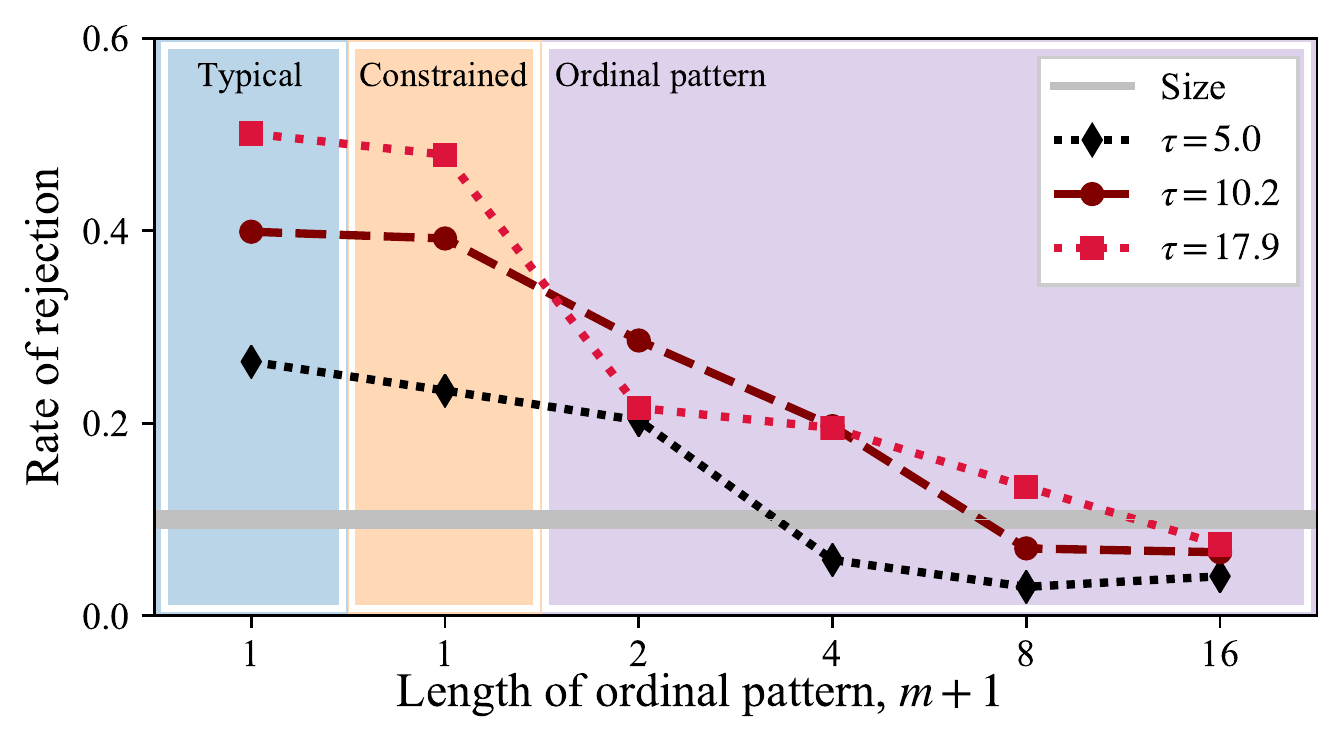}
\caption{By constraining longer ordinal patterns we can accommodate longer correlations in power-law data. Rates of rejection estimated from $1,000$ hypothesis tests with KS-distance as discriminating statistic. Each test uses a sample of length $N = 1,024$ from a power-law which has scale exponent $\alpha = 2.5$ and Lyapunov time $\tau$ equal to $5.0$ (black diamonds with densely dotted lines), $10.2$ (maroon circles with dashed lines) and $17.9$ (red squares with dotted lines) sec/nat (see Appendix \ref{sapp:corr}). Tests use nine typical, constrained, or constrained length $m+1$ ordinal pattern power-law surrogates, have a nominal size 10\%, and use the KS-distance as test statistic. The gray line spans one standard error above and below the nominal size. Error bars show standard error, but are smaller than the marker size.
}
\label{fig:hyp_test_tent_map}
\end{figure}

Next, in Fig. \ref{fig:hyp_test_tent_map}, we use i.i.d. and constrained ordinal pattern power-law surrogates to investigate power-law data of deterministic chaotic origin, having correlation time (Lyapunov time) $\tau = 5.0$, $10.2$, and $17.9$ sec/nat (Appendix~\ref{sapp:corr}). We apply hypothesis tests which employ the discriminating statistic most widely used in tests of a power-law null hypothesis; the KS-distance. As correlation time $\tau$ increases, the considered power-law sequences deviate more from i.i.d., leading to increases in the rejection rates arising from both typical and constrained i.i.d. power-law surrogates. In contrast, constrained ordinal pattern power-law surrogates can accommodate correlations. The length $m + 1$ of ordinal patterns which must be preserved to avoid rates of rejection above the nominal size increases as the correlation time $\tau$ grows, 
showing how ordinal pattern power-law surrogates can be used to resolve correlation structure in observed data. 

\begin{figure}[htbp]
\includegraphics[width=0.45\textwidth]{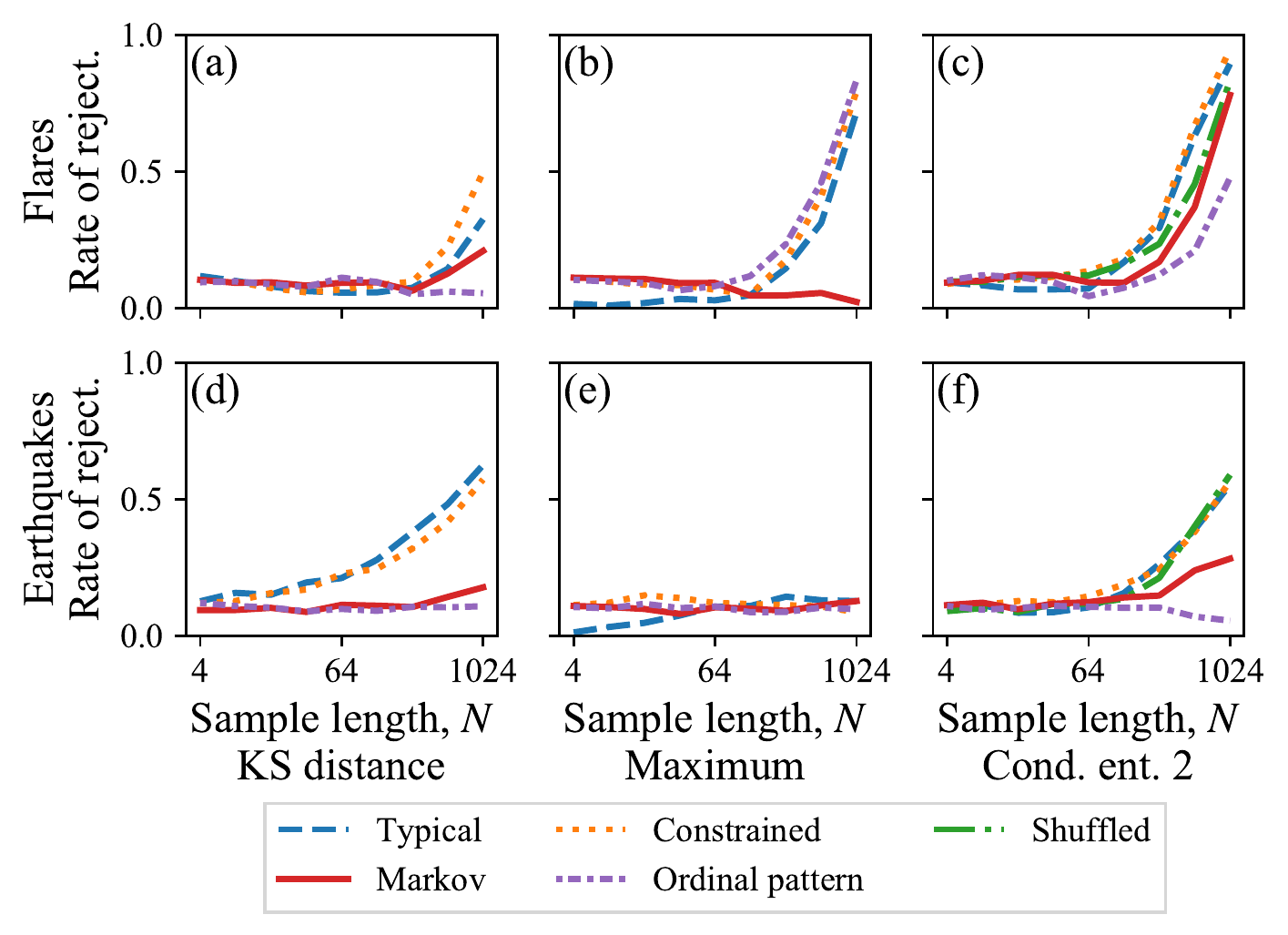}
\caption{Incorporating correlation constraints impacts the rate of rejection of power-law hypotheses for empirical observations. Rates of rejection estimated from $1,000$ hypothesis tests of sequences of length $N$ of (a-c) intensities of solar flares, and (d-f) energy released by earthquakes (Table \ref{tab:real_data_p}). Each input sequence $\xin$ considered begins at an independently and randomly chosen point in the full empirical sequence. Tests use nine typical i.i.d., constrained i.i.d., shuffled, constrained Markov order one or constrained length 16 ordinal pattern surrogates, have nominal size 10\%, and employ as discriminating statistics the KS-distance, maximum or conditional entropy of order two. Note that the conditional entropy of order $m + 1$ is an appropriate statistic for testing for Markov order $m$ (Appendix \ref{sapp:con_ent}).
}
\label{fig:real_corr}
\end{figure}

Finally, in Fig. \ref{fig:real_corr} we illustrate how the choice of surrogate method impacts the conclusions we make about the correlated empirical datasets comprising sequences of solar flare intensities and earthquake energy release. Employing a constrained Markov order one power-law surrogate instead of a typical or constrained i.i.d. power-law surrogate consistently leads to low rates of rejection of a power-law hypothesis on the basis of sample maximum or the KS-distance. However, the type of order one Markov constraints which we consider do not explain the observed order two Markov properties (captured by the conditional entropy of order two) of sequences of solar flare intensities. In contrast, for sequences of earthquake energies these Markov constraints substantially decrease rates of rejection when using the conditional entropy of order two and so appear to be able to at least partly explain the observed memory properties. Constrained length $m + 1 = 16$ ordinal pattern power-law surrogates, which enforce the null hypothesis that sequences are power-law but have correlations which can be captured by ordinal patterns of length $m + 1 = 16$, also lead to low rates of rejection based on the KS-distance (for ordinal patterns of other lengths, other discriminating statistics, and results for another empirical dataset, see SM~\cite{supplemental}, {Fig.~S5,S6}). However, for solar flare intensities, using ordinal pattern surrogates only slightly reduces the rate of rejection when the conditional entropy of order two is used as a discriminating statistic, and actually increases the rate of rejection when using the maximum. When correlations in earthquake intensities are accommodated by constraining ordinal patterns, the discriminating statistics considered consistently lead to low rates of rejection of a power-law model.

\section{Conclusions}

Generating sequences with fat-tailed distributions is a critical step in quantitative investigations of complex systems. The traditional parametric approach, based on fitting a power-law to the data, has the important limitations that it leads to sequences from a single power-law exponent $\alpha$ and ignores correlations in the data. In this paper we proposed non-parametric methods to obtain constrained power-law surrogates which overcome these limitations by not restricting $\alpha$ and by accommodating correlations in the data up to a pre-defined length $m$. We explored the benefit of our surrogates over alternative approaches (shuffling, bootstrapping, and the typical power-law fitting approach) in a variety of settings and datasets. Our approach leads to uniformly distributed p-values, exact hypothesis tests, and unbiased estimates of expectation regardless of the sample statistics. The benefits are particularly important for small sample lengths $N$, which is the most important regime because this is when the determination of the validity of power-law distributions is challenging. This regime is also relevant in large datasets because of the reduction of the effective sample size $N$ that occurs when a power-law is fit only to the tail of a distribution~\cite{clauset2009power,broido2019scale} and when sequences are downsampled to reduce correlations~\cite{gerlach2019testing}.

The significance of our methods and findings is that they provide an improved methodology to address problems recently raised in the long-standing debates on the ubiquity of power-law distributions in complex systems. 
First, it has recently been shown that ignoring correlations, which are ubiquitous in complex systems and affect statistical properties of time series, leads to wrong conclusions about the validity of power-law distributions~\cite{gerlach2019testing}. The usual methods do not account for correlations, but our constrained surrogates do. Second, it is becoming increasingly accepted that whether a distribution is a power-law is less important than whether it has a heavy tail, and that it is important to move beyond goodness-of-fit statistics~\cite{stumpf2012critical,holme2019rare}. Constrained power-law surrogates align with this view because (i) they provide theoretical support for arbitrary discriminating statistics and, hence, a licence to investigate whichever types of deviation from power-law behaviour are most likely to affect the best course of action; and (ii) they provide estimates substantially more accurate than the typical approach when the underlying distribution is log-normal rather than power-law, and we would expect similar advantages to hold for other heavy-tailed distributions. Third, the application of our constrained surrogates modifies conclusions obtained in data analysis, as shown in Fig. \ref{fig:real_corr} above for the case of the analysis of Earthquake datasets; the application of the usual approaches lead to a rejection of the power-law hypothesis (Gutenberg-Richter law) but the incorporation of correlations within our constrained surrogate approach shows that this hypothesis cannot be rejected at a rate significantly higher than the nominal size of the test.

Constrained power-law surrogates can preserve arbitrary additional properties present in the original sequence, but precisely which characteristics should be preserved depends on cases of interest and further work is needed to expand our results to new settings. In the setting of networks, analysis of whether the degree distribution is power-law distributed (scale-free hypothesis) needs to account for constraints related to the formation of networks (e.g., impose that the degree sequence of a simple undirected graph must be graphical~\cite{erdos1960graphs, arratia2005likely} and account for the process of network creation~\cite{chorozoglou2019investigating,falkenberg2020identifying}). It would be interesting to investigate what effect these additional constraints would have on conclusions made about the ubiquity of scale-free networks~\cite{broido2019scale}. In this paper, we focused on discrete power-laws and, before analysis, simply rounded certain originally continuous datasets. Because discretization and rounding are common steps in data collection, interpretation and analysis~\cite{serinaldi2017general}, it would be valuable systematically to study the impact these have on conclusions and estimates about (potentially) fat-tailed datasets. Relationships which are theoretically assured for discrete data might hold only approximately for continuous power-laws. For this reason, and also because the additional analytic properties of real continuous data could provide new opportunities~\cite{diks1995reversibility, kugiumtzis1999test, kugiumtzis2002statically,dimitriadis2018stochastic, lancaster2018surrogate}, it would be valuable separately to develop constrained continuous power-law surrogates which can be used to explore unobserved events.

\begin{acknowledgements}
JMM and GY are supported by National Natural Science Foundation of China (Grant No. 11875043{, 12150410309}), Shanghai Municipal Science and Technology Major Project (Grant No. 2021SHZDZX0100), Shanghai Municipal Commission of Science and Technology Project (Grant No. 19511132101) and the Fundamental Research Funds for the Central Universities (Grant No. 22120190251). We thank Michael Small, Thomas J\"ungling, D\'ebora Cristina Corr\^ea, Yinqi Xuan and anonymous reviewers for valuable input.
\end{acknowledgements}

\appendix

\section{Correlated power-law sequences} \label{sapp:corr}

Here we describe how we generate input sequences $\xin$ which are power-law distributed, but not i.i.d. We consider two cases: (1) a stochastic generative process of Markov order $m > 0$; and (2) a deterministic chaotic system with tuneable correlation time $\tau$.

\subsection{Stochastic process}

Sequences from power-laws of Markov order $m > 0$ are obtained using as Markov states the partition $\mathcal{A} = \left\{ a_i \mid i \in \mathbb{Z}^+ \right\}$ of the set of integers greater than or equal to the lower cut-off $x_{\min}$ which comprises intervals of equal logarithmic width defined by $a_i = \left\{k \in \mathbb{Z} \mid 3^{i-1} \left(x_{\min} - 0.5\right) \leq k < 3^i \left(x_{\min} - 0.5\right)\right\}$. The first $m$ elements $x_1, \ldots, x_m$ of the sequence are generated i.i.d. from the limiting power-law distribution. For each integer time $t > m$, with probability $\nu$ the element $x_t$ is also generated i.i.d. from the limiting power-law distribution. Otherwise, the element $x_t$ is instead generated by first choosing one of the previous $m$ values of $x$ (say, $x_*$), identifying the interval $a_{i_{*}}$ such that $x_* \in a_{i_{*}}$, and choosing the value $x_t$ randomly within $a_{i_{*}}$ according to the corresponding power-law distribution:
\begin{align}
p\left(x\mid x \in a_{i_{*}}\right)
&= \begin{cases}
x^{-\alpha}/\sum\limits_{y \in a_{i_{*}}} y^{-\alpha}, & x \in a_{i_*}\\
0, & x \notin a_{i_*}
\end{cases}.\notag
\end{align}

We now check that this algorithm leads to the desired limiting distribution in the case of Markov of order one (the same reasoning applies for higher order). The probability $p_{t + 1}(a)$ of Markov state $a \in \mathcal{A}$ at time $t + 1$ given the probability $p_t(a)$ at time $t$ is
\begin{equation}
    p_{t+1}(a) = \nu p_{\text{limiting}}(a) + (1 - \nu) p_t(a),\label{eq.corr}
\end{equation}
where $p_{\text{limiting}}$ is the desired limiting distribution of Markov states and $\nu$ is a parameter governing the probability of choosing the next value according to $p_{\text{limiting}}$. Iterating Eq. (\ref{eq.corr}) we find that, for $a \in \mathcal{A}$ and non-negative integer $t$,

\begin{equation}
p_t(a) = \nu \left(\sum\limits_{j=0}^t (1 - \nu)^j\right) p_{\text{limiting}}(a) + (1 - \nu)^t p_0(a)\notag
\end{equation}
and so $\lim\limits_{t\rightarrow\infty} p_t(a) = p_{\text{limiting}}(a)$ for any $0<\nu\le 1$. In our case, $p_{\text{limiting}}$ is a power-law with scale exponent $\alpha = 2.5$ and lower cut-off $x_{\min} = 1$, and $\nu = 0.1$.

\subsection{Deterministic chaotic system}

We generate power-law sequences with tuneable correlation time $\tau$ using a one-dimensional deterministic chaotic system. This is achieved by first generating a sequence from a map with a tuneable correlation (Lyapunov time) and then applying a transformation of variables that maps the invariant density to a power-law distribution. We use the asymmetric tent map $f:[0, 1] \rightarrow [0, 1]$ given by
\begin{align}
w_{t+1} = f\left( w_t \right) = \begin{cases}
r^{-1} w_t, & 0 \leq w_t \leq r\\
(1 - r)^{-1} \left( 1 - w_t \right), & r < w_t \leq 1
\end{cases},\notag
\end{align}
where $r \in [0, 1]$ is a parameter, and iterate starting from a point $w_1 \in [0,1]$ chosen uniformly at random  to produce a sequence ${\bf w} = w_1, \ldots, w_N$ of length $N$. This map is uniformly hyperbolic with a Lyapunov exponent given by ~\cite{van1997chaos}
\begin{align}
\lambda = \left\lvert r \log r + (1 - r) \log (1 - r)\right\rvert.\notag
\end{align}
The inverse of the Lyapunov exponent $\tau = 1/\lambda$ provides a measure of the correlation (memory) in the sequence. We consider parameters $r = 0.95, 0.98$ and $0.99$, corresponding to $\tau = 5.0, 10.2$ and $17.9$ sec/nat respectively (the Lyapunov exponent has units of nat/sec because we use natural logarithms).
The invariant density of the map is uniform in $[0,1]$~\cite{gaspard2005chaos}, which allows us to obtain a correlated power-law sequence ${\bf x} = x_1, \ldots, x_N$. This is done mapping each value $w_t \in [0, 1]$ to an integer $x_t$ no smaller than the desired lower cut-off $x_{\min}$ via the inverse of the cumulative distribution function of a discrete power-law with the desired scale exponent $\alpha$.  

\section{Surrogate algorithm} \label{sapp:details_alg}

In this appendix we detail our method for producing constrained power-law surrogates with arbitrary lower cut-off $x_{\min} \in \mathbb{Z}^+$ (the code is available in Ref.~\cite{github}). Given an input sequence $\xin = x_1, x_2, \ldots, x_N$, let $x_t = \prod\limits_{j=1}^{{r_t}} q_{t, j}$ be the prime decomposition of element $x_t$, where $q_{t, 1} \geq q_{t, 2} \geq \ldots \geq q_{t, {r_t}}$ and ${r_t}$ is the total number of instances of prime factors in {$x_t$}. As mentioned in the main text, our strategy involves three steps: 
\begin{itemize}
    \item[(1)] Associate with each element $x_t = \prod\limits_{j=1}^{r_t} q_{t, j}$ the $k_t$ prime numbers $q_{t, 1}, q_{t, 2}, \ldots, q_{t, k_t}$, where $k_t$ is chosen such that $\prod\limits_{j = 1}^{k_t} q_{t, j} \geq x_{\min}$ but, for $l = 1, 2, \ldots,  k_t - 1$, $\prod\limits_{j = 1}^l {q_{t, j}} < x_{\min}$. 
    Our strategy involves keeping these $k_t$ instances of prime factors with element $x_t$, and thus ensuring that $x_t$ does not decrease beneath the lower cut-off $x_{\min}$ or, indeed, beneath its personal lower cut-off $x_{t, \min} = \prod\limits_{j = 1}^{k_t} q_{t, j} \geq x_{\min}$. For each element $x_t$, note the maximum admissible prime factor $q_{t, k_t}$.
    \item[(2)] Randomly allocate all instances of prime factors which are not associated with an element of the sequence. For each distinct prime $q$, all instances of this prime factor which are not associated with a particular element of the input sequence $\xin$ are distributed among all elements of the sequence for which the maximum admissible factor is no smaller than $q$ (i.e., all elements $x_t$ for which $q_{t, k_t} \geq q$). So that surrogates are drawn uniformly at random from a set of sequences which have the same product and the same likelihood, the instances of the prime factor $q$ are distributed such that all distinct redistributions are equally likely. Because an element $x_t$ will receive no additional instances of prime factors $q > q_{t, k_t}$, its personal lower cut-off $x_{t, \min} = \prod\limits_{j = 1}^{k_t} q_{t, j}$ will not change. Furthermore, each distinct time series with the same likelihood and the same sequence of personal lower cut-offs $x_{1, \min}, x_{2, \min}, \ldots, x_{N, \min}$ is equally likely. 
    \item[(3)] Randomly permute the time series so that more distinct sequences are accessible. Whether this permutation is performed in the first or final stage of the surrogate algorithm does not affect the probability distribution of sequences. 
\end{itemize}
The algorithm just described involves ordering prime factors from largest to smallest. Any other ordering of the prime numbers would be valid, but would lead to a distinct surrogate algorithm. Our choice preferentially fixes large prime factors and allows smaller factors (which we expect to tend to be more numerous) to be redistributed.

Choosing uniformly at random from all distributions of $k$ instances of a prime factor among $N$ elements is equivalent to choosing a random weak composition of $k$ into $N$ parts, for which we use Algorithm RANCOM~\cite{nijenhuis2014combinatorial}. We follow Ref.~\cite{stumpf2017combalg} in choosing a random $(N - 1)$-subset using the $O(N)$ Algorithm RKS2 instead of the $O(N \log N)$ Algorithm RKSB originally employed as a subroutine of Algorithm RANCOM~\cite{nijenhuis2014combinatorial}. The computational times required for constrained and typical i.i.d. power-law surrogates {both scale almost linearly} in sample length $N$ (see SM~\cite{supplemental}, {Fig.~S7}).

When either constructing constrained i.i.d. power-law surrogates with $x_{\min} \geq 2$ or generating constrained correlated power-law surrogates, we would expect our methods to benefit from the presence of more instances of prime factors, because this should allow more randomization.

{As stated in the main text, the lower cut-off $x_{\min}$ describing an empirical data set of length $N$ is chosen as the value $\hat{x}_{\min}$ which minimises the KS-distance between the empirical distribution for the $N_G$ values above the lower cut-off and the corresponding maximum likelihood discrete power-law on $x \geq x_{\min}$~\cite{bauke2007parameter,clauset2009power,broido2019scale}. This fitted lower cut-off is actually treated in three slightly different ways depending on the surrogate methods considered:
{(1) We fit the lower cut-off $x_{\min}$ only once for each empirical dataset, using all available data. Thereafter we work with (subsamples or subsequences of) the sequence of elements no less than this fitted lower cut-off $\hat{x}_{\min}$, and treat the previously fitted value $\hat{x}_{\min}$ as the known value of the lower cut-off $x_{\min}$. In particular, the KS-distance is calculated using this previously fitted $\hat{x}_{\min}$. This approach has been applied in all cases, except in Table~\ref{tab:real_data_p}. 
}
(2) Following Ref.~\cite{clauset2009power}, each typical surrogate considered in Table~\ref{tab:real_data_p} comprises a sequence of $N$ values chosen i.i.d., as follows. With probability $N_G/N$, the value is drawn i.i.d. from the maximum likelihood power-law for the $N_G$ values no less than the fitted lower cut-off $\hat{x}_{\min}$. Otherwise (i.e., with probability probability $\left( N- N_G\right)/N$), the value is drawn i.i.d. from the $\left( N - N_G\right)$ elements which fall below the fitted lower cut-off $x_{\min}$. The KS-distance is calculated only after re-fitting the lower cut-off to the surrogate sequence~\cite{clauset2009power}. (3) Each constrained surrogate of Table~\ref{tab:real_data_p} is generated while constraining the fitted lower cut-off $\hat{x}_{\min}$ as well as the maximum likelihood value $\hat{\alpha}$ of the scale exponent $\alpha$. Each such} surrogate comprises the values in the original observation which fell below the fitted lower cut-off $\hat{x}_{\min}$, together with a constrained power-law surrogate generated from the observations with values above the fitted lower cut-off. {To constrain the fitted lower cut-off $\hat{x}_{\min}$}, we accept a surrogate only when it leads to the same fitted value of $\hat{x}_{\min}$ as does the original observation, but in other cases discard the result and repeat the surrogate generation process. {Conditioning} on the fitted lower cut-off $\hat{x}_{\min}$ increases computational cost, {which is why we consider this approach only in Table~\ref{tab:real_data_p}. 
} 


{
\section{Implementing hypothesis tests} \label{sapp:imp_hyp}\label{sapp:con_ent}
In this appendix we detail the implementation of hypothesis tests for an observed real sequence $\xin = x_1, \ldots, x_N$ and a given null hypothesis. Before beginning a test, we choose: (1) a discriminating statistic $s:\mathbb{R}^N \rightarrow \mathbb{R}$ (in the main text we use the KS-distance, the mean, the variance, the maximum, and conditional entropies of order one and two); (2) a size $\alpha_{\text{size}} \in (0, 1)$, which corresponds to the maximum allowable probability of incorrectly rejecting the null hypothesis even when it is true, also called the rate of false positives or Type I error (in the manuscript we use $\alpha_{\text{size}}=0.1$); (3) a number $N_s$ of surrogates ${\bf x}$ independently to generate (in the main text we use $N_s = 9$, except in Table~\ref{tab:real_data_p}, where we use $N_s = 999$); and (4) whether the test should be left-tailed (appropriate when lower values of the discriminating statistic are more extreme/less consistent with the null hypothesis), right-tailed (appropriate when higher values of the discriminating statistic are more extreme/less consistent with the null hypothesis), or two tailed (appropriate when both unexpectedly low and unexpectedly high values of the discriminating statistic correspond to noteworthy evidence against the null hypothesis). 
A quantile $q$ is estimated as
\begin{align}
    q = \frac{r - 0.5}{N_s + 1} \in (0, 1),\notag
\end{align}
where the rank $r \in \left\{1, 2, \ldots, N_s + 1\right\}$ is the position of the value $\sinput = s \left( \xin \right)$ of the discriminating statistic observed for the input sequence $\xin$ when it is combined with the $N_s$ values $\left\{s_n = s \left( {\bf x}_n\right)\right\}$ observed for the $N_s$ surrogate sequences ${\bf x}_n$, and these $N_s + 1$ values are ranked from smallest to largest, with ties broken by adding small random perturbations. The p-value $p \in (0, 1)$ can then be computed from the quantile $q$ according to
\begin{align}
    p = 
    \begin{cases}
    q & \text{if the test is left-tailed,}\\
    1 - q & \text{if the test is right-tailed,}\\
    2 \min \{q, 1 - q\} & \text{if the test is two-tailed}.
    \end{cases}
    \notag
\end{align}
Finally, if $p < \alpha_{\text{size}}$ ($p \geq \alpha_{\text{size}}$) then we reject (fail to reject) the null hypothesis at the level of significance $\alpha_{\text{size}}$. 
}

{
The most popular method to assess a power-law hypothesis involves a right-tailed test using the KS-distance as the discriminating statistic~\cite{clauset2009power,broido2019scale}. 
The best combination of statistic and tail depends on the specific type of deviations of most interest or practical importance. We have used left-tailed tests together with the mean, maximum and variance because power-law distributions are often used to investigate extreme events or to represent fat-tailed processes. We would intuitively expect lower values for these discriminating statistics to correspond to lower risk from extreme events, less support for the hypothesis of a fat-tailed process, and less motivation to model these processes using power-law distributions. }

A test of the null hypothesis that a sequence arose from a Markov chain of order $m$ should use a statistic which is sensitive to Markov properties of order greater than $m$, because this allows detection of differences between surrogates of Markov order $m$ and original data of higher Markov order. Such a statistic is an estimate of the conditional entropy $h_{m + 1}$ of order $m + 1$, which for a stationary sequence $z_1, z_2, \ldots, z_N$ is given by
\begin{align}
h_{m+1} = H_{m+1} - H_m,
\notag
\end{align}%
where $H_m$ is the entropy of order $m$
\begin{align}
    H_m = - \sum\limits_{z_{t}, z_{t + 1}, \ldots, z_{t + m}} &p (z_{t}, z_{t + 1}, \ldots, z_{t + m})\notag\\
    &\times \log p (z_{t}, z_{t + 1}, \ldots, z_{t + m}),\notag
\end{align}
and $p (z_{t}, z_{t + 1}, \ldots, z_{t + m})$ is the probability that a subsequence of length $m + 1$ will be $z_{t}, z_{t + 1}, \ldots, z_{t + m}$. By stationarity, this probability is independent of $t$. Because we are interested in both the power-law and Markov properties of sequences, we consider conditional entropies defined in terms of the original time series $x_1, x_2, \ldots, x_N$ rather than the sequence of Markov states $z_1, z_2, \ldots, z_N$. Joint probabilities are estimated from the observed sequence using the maximum-likelihood method, for which probabilities are proportional to the frequency. We use natural logarithms, so that entropies have units of nats.
{Following, e.g., Ref.~\cite{van1998testing,pethel2014exact,correa2020constrained}, tests involving conditional entropies are also left-tailed.}

\bibliography{ref}
\end{document}


\makeatletter
\@fpsep\textheight
\makeatother

\interfootnotelinepenalty=10000

\renewcommand{\theequation}{S\arabic{equation}}\renewcommand{\thefigure}{S\arabic{figure}}
\makeatletter
\def\@biblabel#1{[S#1]}
\makeatother
\newcommand{\scite}[1]{\textrm{[S\citealp{#1}]}}


\preprint{APS/123-QED}

\title{Supplemental material: Non-parametric power-law surrogates}

\author{Jack Murdoch Moore}
\author{Gang Yan}%
\author{Eduardo G. Altmann}%


\begin{abstract}
\end{abstract}

\maketitle




\section{Generating constrained surrogates by fixing likelihood}

In this section we demonstrate properties of constrained surrogates chosen uniformly at random from a set $\mathcal{U} \left( \xin \right)$ satisfying conditions C1-C3 of Sec.~II~A of the main paper, focusing on the case in which the likelihood under the null hypothesis is one of the constrained properties. We show that, under the null hypothesis, this prescription is equivalent to conditioning on the outcome of the map $\mathcal{U} : \xin \mapsto \mathcal{U} \left( \xin \right)$, and that the corresponding conditional probability is independent of model parameters. 


\noindent{\bf Proof.} 

Let ${\bf X}$ be a discrete random variable with domain $\mathcal{D}$ and probability distribution $P_{\alpha}\left({\bf x}\right)$, where $\alpha$ denotes the model parameter(s). Let $S = \mathcal{U} \left( {\bf X} \right)$, and let $P_{\alpha}\left({\bf x} \mid s\right)$ denote the conditional probability of ${\bf X}$ given $S$. We will show that for any value of the model parameter $\alpha$, and for any ${\bf x} \in \mathcal{D}$,
\begin{equation}
P_{\alpha}\left({\bf x} \mid s\right) = \begin{cases}{\left\lvert \mathcal{U}\left({\bf x}\right)\right\rvert}^{-1},&s = \mathcal{U}\left({\bf x}\right)\\
0,&s \neq \mathcal{U}\left({\bf x}\right)
\end{cases},\notag
\end{equation}
where $\left\lvert \mathcal{A} \right\rvert$ denotes the cardinality of a set $\mathcal{A}$. This expression for the conditional probability $P_{\alpha}\left({\bf x} \mid s\right)$ is independent of $\alpha$. In addition, the probability corresponds to choosing uniformly at random from the set $\mathcal{U}\left({\bf x}\right)$, showing that, under the null hypothesis, the way in which we produce constrained surrogates is indeed equivalent to conditioning on $\mathcal{U}$.

Let $Q_{\alpha} \left(s\right)$ denote the probability distribution of $S = \mathcal{U}\left({\bf X}\right)$, and let $P_{\alpha}({\bf x}, s)$ denote the joint distribution of ${\bf X}, S$. Let ${\bf x},{\bf y} \in \mathcal{D}$. If $\mathcal{U}\left({\bf x}\right) \neq \mathcal{U}\left({\bf y}\right)$ then $P_{\alpha}\left({\bf x} \mid \mathcal{U}\left({\bf y}\right)\right) = 0$. Hence, we can restrict our attention to the case
\begin{equation}\label{eq.joint}
\mathcal{U}\left({\bf x}\right) = \mathcal{U}\left({\bf y}\right),
\end{equation}%
in which circumstance
\begin{align}
P_{\alpha}\left( {\bf x}, \mathcal{U}\left({\bf y}\right) \right) &= P_{\alpha}({\bf x}, \mathcal{U}\left({\bf x}\right))\notag\\
&= P_{\alpha}\left({\bf x}\right).\label{eq.characteristic}
\end{align}%
First we note that, because the likelihood given by $\mathcal{L}_{{\bf x}}\left( \alpha \right) = P_{\alpha} \left( {\bf x}\right)$ is one of the constrained properties $K$, condition C3 implies
\begin{align}
\forall {\bf z} \in \mathcal{U}\left({\bf x}\right),\qquad P_{\alpha}\left( {\bf z} \right) &= P_{\alpha}\left( {\bf x} \right).\label{eq.prob}
\end{align}%
Now we will show that, by conditions C1 and C2,
\begin{align}\label{eq.partition}
\forall {\bf z} \in \mathcal{D},\qquad \mathcal{U}\left({\bf z}\right) = \mathcal{U}\left({\bf x}\right) \Leftrightarrow {\bf z} \in \mathcal{U}\left({\bf x}\right).
\end{align}%
($\Rightarrow$): For any ${\bf z} \in \mathcal{D}$, if $\mathcal{U}\left({\bf z}\right) = \mathcal{U}\left({\bf x}\right)$ then, by condition C1, ${\bf z} \in \mathcal{U}\left({\bf z}\right) = \mathcal{U}\left({\bf x}\right)$. ($\Leftarrow$): Conversely, for any ${\bf z} \in \mathcal{D}$, if ${\bf z} \in \mathcal{U}\left({\bf x}\right)$ then, by condition C2, $\mathcal{U}\left({\bf z}\right) = \mathcal{U}\left({\bf x}\right)$. Result (\ref{eq.partition}) implies that
\begin{align}
Q_{\alpha}(\mathcal{U}\left({\bf x}\right)) &= \sum_{\substack{{\bf z} \in \mathcal{D}\\ \mathcal{U}\left({\bf z}\right) = \mathcal{U}\left({\bf x}\right)}} P_{\alpha}\left({\bf z}\right)\notag\\
&= \sum_{{\bf z} \in \mathcal{U}\left({\bf x}\right)} P_{\alpha}\left({\bf z}\right),\notag
\end{align}%
showing, together with Eq.~(\ref{eq.prob}), that
\begin{align}
Q_{\alpha}(\mathcal{U}\left({\bf x}\right)) &= \sum_{{\bf z} \in \mathcal{U}\left({\bf x}\right)} P_{\alpha}\left({\bf x}\right)\notag\\
&= \left\lvert\mathcal{U}\left({\bf x}\right)\right\rvert P_{\alpha}\left({\bf x}\right).\label{eq.uniform}
\end{align}
If $P_{\alpha}\left({\bf x}\right) \neq 0$,
\begin{align}
P_{\alpha}\left({\bf x} \mid \mathcal{U}\left({\bf y}\right)\right) &= \frac{P_{\alpha}\left({\bf x}, \mathcal{U}\left({\bf y}\right)\right)}{Q_{\alpha}(\mathcal{U}\left({\bf y}\right))}\notag\\&= \frac{P_{\alpha}\left({\bf x}\right)}{Q_{\alpha}(\mathcal{U}\left({\bf x}\right))}&&\textrm{By Eq. (\ref{eq.joint}, \ref{eq.characteristic})}\notag\\
&= \frac{P_{\alpha}\left({\bf x}\right)}{\left\lvert\mathcal{U}\left({\bf x}\right)\right\rvert P_{\alpha}\left({\bf x}\right)}&&\textrm{By Eq. (\ref{eq.uniform})}\notag\\
&= {\left\lvert\mathcal{U}\left({\bf x}\right)\right\rvert}^{-1}.\notag
\end{align}%
We can neglect the case $P_{\alpha}\left({\bf x}\right) = 0$ because in such a situation, by Eq. (\ref{eq.uniform}), $Q_{\alpha}(\mathcal{U}\left({\bf x}\right)) = 0$ and the conditional probability $P_{\alpha}\left({\bf x} \mid \mathcal{U}\left({\bf y}\right)\right)$ is not defined.
$\square$



\begin{figure*}[p]
\centering
\includegraphics[width=0.6\textwidth]{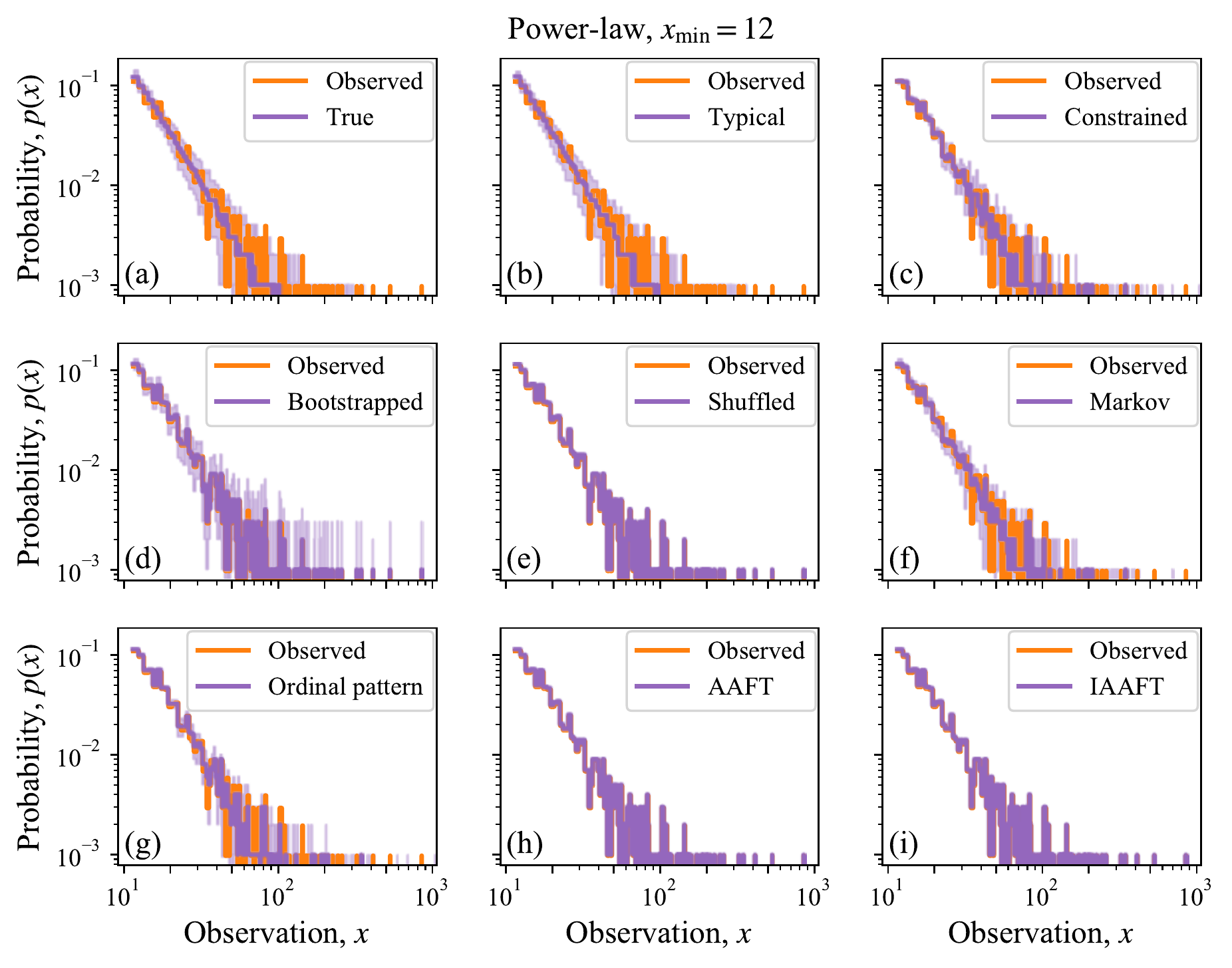}
\caption{
{Bootstrap, shuffle, AAFT, and IAAFT surrogates preserve power-law characteristics but cannot generate previously unobserved values; typical and constrained power-law surrogates preserve statistical characteristics of input power-law data while randomising specific values.}
Each panel shows: (1) the probability distribution of an input sequence $\xin$ arising from $N = 1024$ i.i.d. power-law random variables with scale exponent $\alpha = 2.5$ and lower cut-off $x_{\min} = 12$ (opaque orange line); (2) the median probability over 100 independent surrogate realisations from this input sequence $\xin$ (opaque purple line); and (3) the 90\% confidence interval of the probability (transparent purple region). The surrogates considered are: (a) the true underlying process (i.i.d. power-law with scale exponent $\alpha = 2.5$ and lower cut-off $x_{\min} = 12$); (b) typical i.i.d. power-law; (c) constrained i.i.d. power-law; (d) bootstrap; (e) shuffle; (f) constrained Markov order one power-law; (g) constrained length $m + 1 = 16$ ordinal pattern power-law; (h) amplitude adjusted Fourier transform (AAFT); and (i) iterated amplitude adjusted Fourier transform (IAAFT).
}
\label{fig:histograms}
\end{figure*}


\begin{figure*}[p]
\centering
\includegraphics[width=0.80\textwidth]{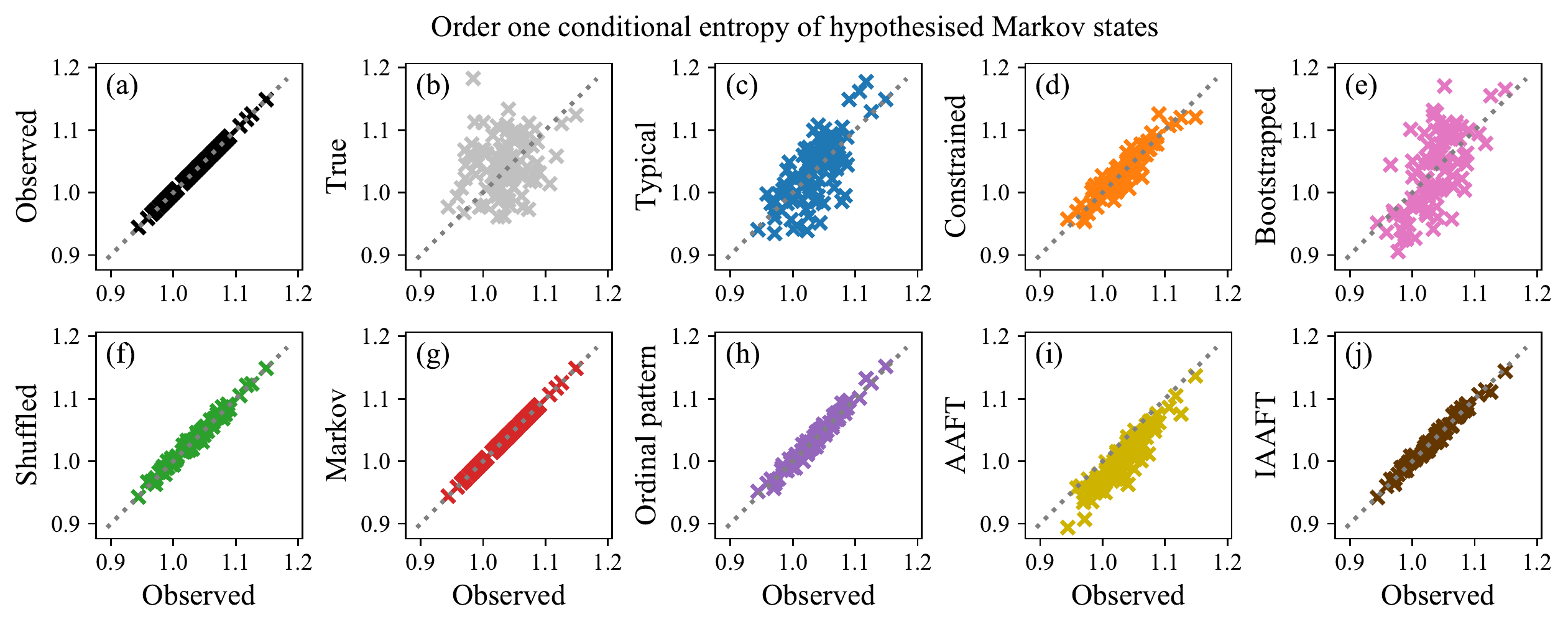}
\caption{
{Among all randomizing surrogates, only the} Markov order one power-law surrogate preserves the conditional entropy of order one of the hypothesised Markov states. Each panel shows 100 markers, each of which corresponds to an independently generated observed  input sequence $\xin$ comprising $N = 1024$ independently and identically distributed (i.i.d.) power-law random variables with scale exponent $\alpha = 2.5$ and lower cut-off $x_{\min} = 1$. The $x$-coordinate ($y$-coordinate) of each marker is the conditional entropy of order one of hypothesised Markov states of the corresponding observed input sequence (a single surrogate generated from the corresponding observed input sequence). The surrogates considered are: (a) exact reproduction of the observed sequence (trivially producing markers along the line of identity $y = x$); (b) the true underlying process (i.i.d. power-law with scale exponent $\alpha = 2.5$); (c) typical i.i.d. power-law; (d) constrained i.i.d. power-law; (e) bootstrap; (f) shuffle; (g) constrained Markov order one power-law; (h) constrained ordinal pattern power-law; (i) amplitude adjusted Fourier transform (AAFT); and (j) iterated amplitude adjusted Fourier transform (IAAFT). {The conditional entropy considered is relative to the Markov states, based on log-binning (main paper, Appendix A), defining the constrained Markov order one power-law surrogate algorithm which we employ.}
}
\label{fig:cond_ent_one}
\end{figure*}

\begin{figure*}[p]
\includegraphics[width=0.50\textwidth]{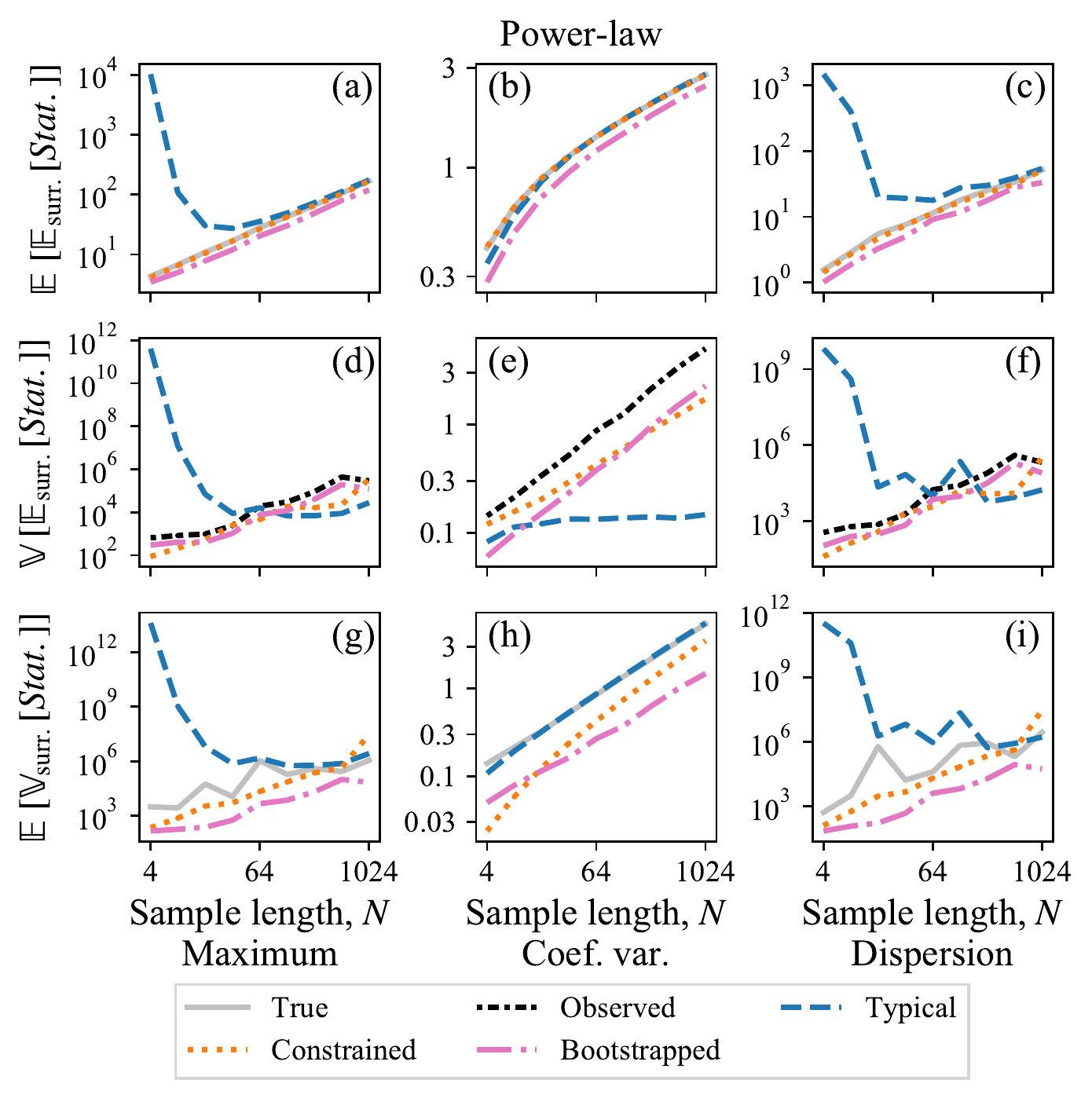}
\caption{
Constrained surrogates do not bias sample statistics and can reduce their variance. (a-c) Expectation $\mathbb{E}$ and (d-f) variance $\mathbb{V}$ of estimated expectation $\mathbb{E}_{\text{surr.}}$ of sample statistics. (g-i) Expectation $\mathbb{E}$ of estimated variance $\mathbb{V}_{\text{surr.}}$ of sample statistics.
$\mathbb{E}$ and $\mathbb{V}$ are calculated using $10,000$ independent input sequences $\xin$, each comprising $N$ values drawn i.i.d. from a power-law distribution with $\alpha = 2.5$.
$\mathbb{E}_{\text{surr.}}$ and $\mathbb{V}_{\text{surr.}}$ are estimated using $100$ surrogates, each from the same $\xin$.  The surrogates considered are either generated independently from the true underlying process (solid gray), exactly the input sequence $\xin$ originally observed (dot-dashed black), typical, constrained or bootstrapped. Surrogates have the same length $N$ as the original sequence (horizontal axis). Results for the sample maximum, coefficient of variation, and index of dispersion are shown in each column and {are} computed in sequences of length $N$. 
{Constrained surrogates provide an unbiased estimate in all cases, in contrast to bootstrapping and the typical approach.
As predicted by Eq.~(4) of the main paper, taking the mean $\mathbb{E}_{\text{surr.}}$ over many constrained surrogates reduces finite variance $\mathbb{V}$ in estimates of the expectation of the sample coefficient of variation. Because the sample maximum and sample index of dispersion of a power-law with scale exponent three or less have infinite variance, taking the mean $\mathbb{E}_{\text{surr.}}$ over many constrained surrogate realisations may not reduce its variance $\mathbb{V}$, although it usually reduces the variance we estimate from a finite number of independently realised power-law sequences $\xin$. Typical surrogates can also reduce variance $\mathbb{V}$, but may introduce bias at the same time. 
The variance $\mathbb{V}_{\text{surr.}}$ of a statistic over many typical surrogate realisations can be similar to the total variance of that statistic. As expected from Eq.~(4) of the main paper, for finite variance statistics, the variance $\mathbb{V}_{\text{surr.}}$ over constrained surrogates is lower in expectation than the total statistical variance.
The variances $\mathbb{V}_{\text{surr.}}$ and $\mathbb{V}$ of these statistics tend to increase as sample length $N$ increases. This occurs because the sample maximum, coefficient of variation and index of dispersion increase with sample length $N$, but the number of instances of sequences used to calculate variance does not change.
}
}
\label{fig:mean_var_statistics}
\end{figure*}

\begin{figure*}[p]
\centering
\includegraphics[width=0.90\textwidth]{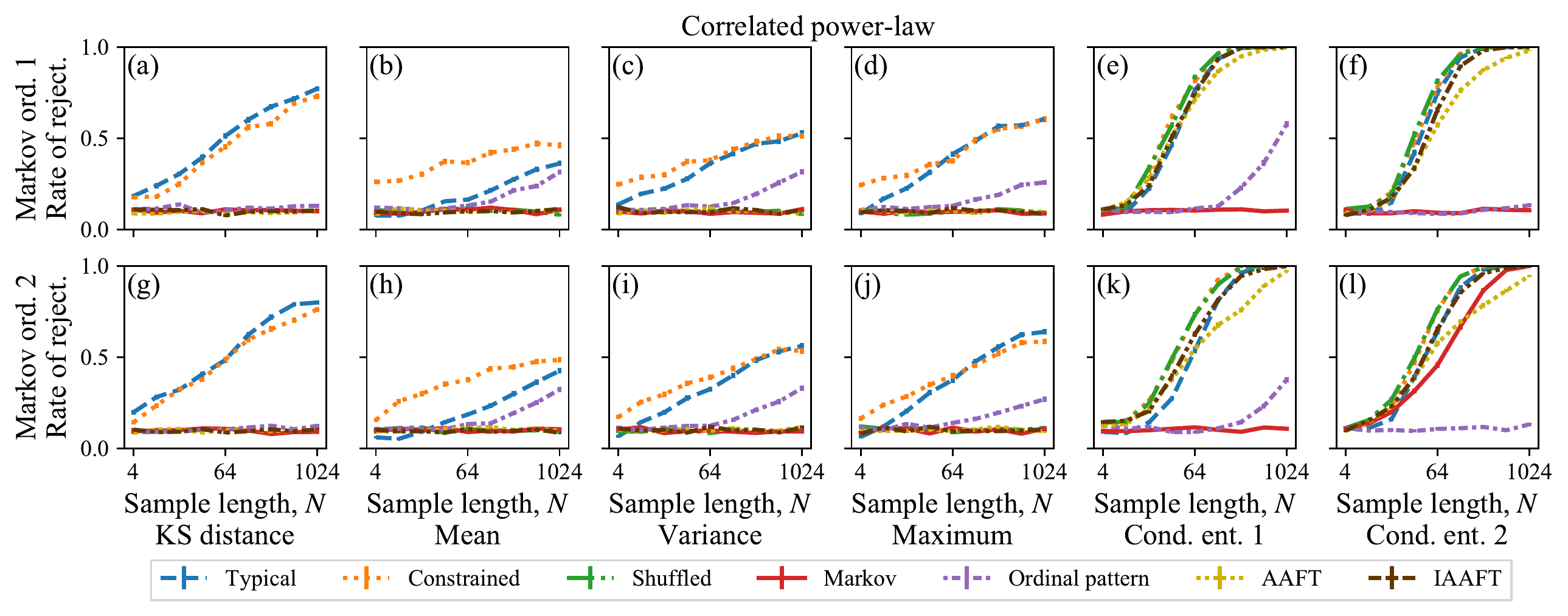}
\caption{
Constrained correlated power-law surrogates reduce correlation-based rejection of power-law hypotheses. 
Rates of rejection estimated from $1,000$ hypothesis tests, each using a sample of length $N$ from a power-law which has scale exponent $\alpha = 2.5$, lower cut-off $x_{\min} = 1$ and Markov order (a-f) one or (g-l) two. Tests use nine typical {i.i.d.} power-law, constrained {i.i.d.} power-law, shuffled, constrained Markov order one power-law, constrained length $m + 1 = 16$ ordinal pattern power-law, AAFT or IAAFT surrogates, have a nominal size 10\%, and use as test statistic (from left to right) the KS distance, mean, variance, maximum, conditional entropy of order one, and conditional entropy of order two. Markov power-law surrogates use the same Markov states as the original time series. Error bars show standard error and are at most only slightly larger than the line width. 
{When shuffle, AAFT or IAAFT surrogates (which only alter the order of a sequence, but not the frequencies of distinct values) are used together with KS distance, mean, variance or maximum (statistics independent of the order), the rate of rejection trivially remains close to the nominal size. When a conditional entropy is used as discriminating statistic together with AAFT or IAAFT surrogates, rejection rates are uniformly high for long samples. Constrained ordinal pattern power-law surrogates do not reject at a rate substantially higher than the nominal size when either the KS distance or the conditional entropy of order two is used as a discriminating statistic. When the mean, variance, maximum or conditional entropy of order one is used instead, constrained ordinal pattern surrogates lead to a rate of rejection lower than that observed for typical or constrained i.i.d. power-law surrogates.}
}
\label{fig:hyp_test}
\end{figure*}

\begin{figure*}[p]
\centering
\includegraphics[width=0.90\textwidth]{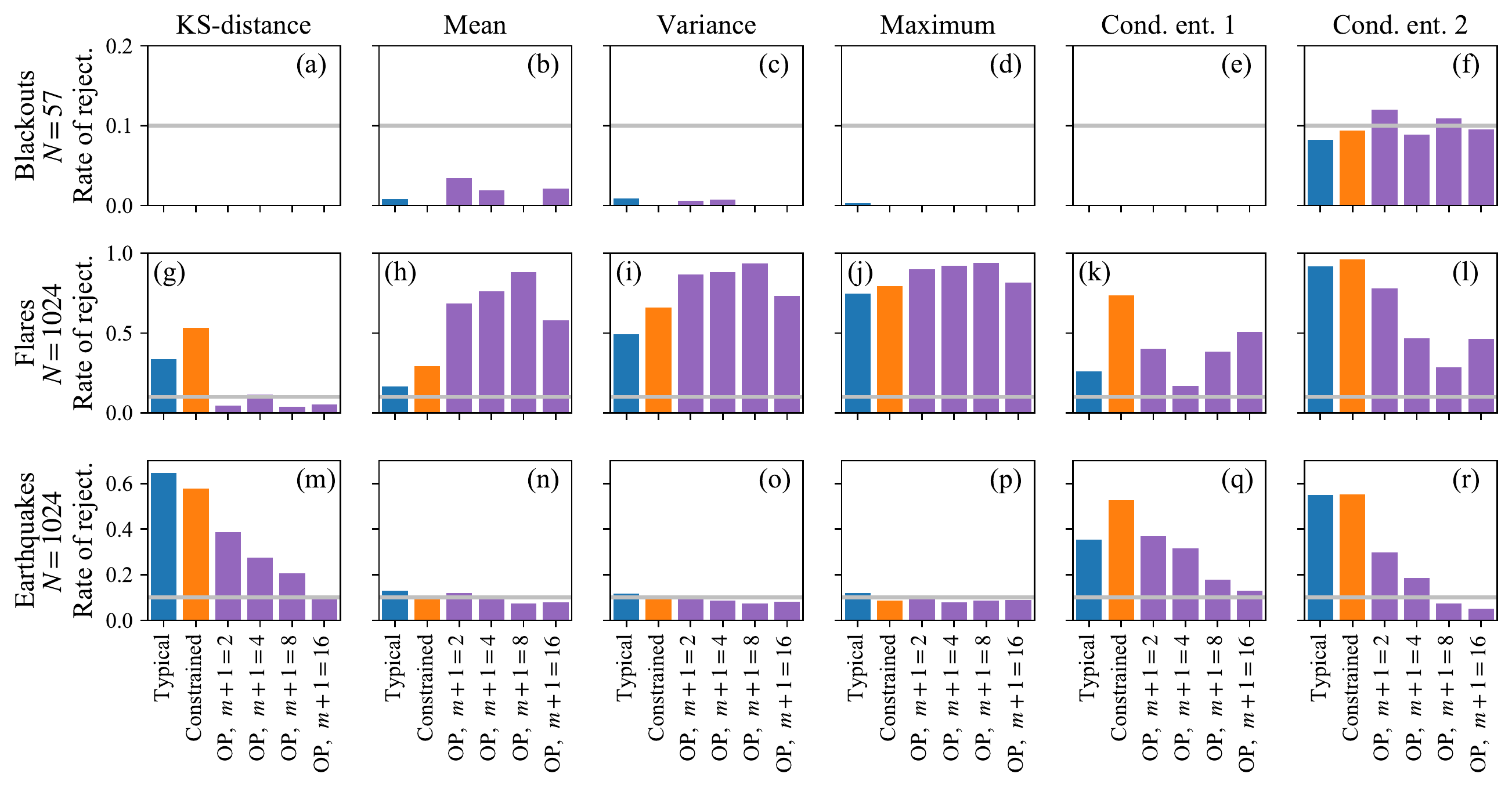}
\caption{For sequences of energy released by earthquakes, but not intensity of solar flares, {constraining ordinal patterns of length $m + 1 = 16$ avoids} rejection of a power-law hypothesis at a rate {substantially} higher than the nominal size. Rates of rejection estimated from $1,000$ hypothesis tests, each using a sample of length (a-f) {$N = 57$ of} number of customers affected by earthquakes, (g-l) {$N = 1,024$ of} intensities of solar flares, and (m-r) {$N = 1,024$ of} energy released by earthquakes. Each input sequence $\xin$ considered begins at an independently and randomly chosen point in the full empirical sequence of values above the fitted lower cut-off $\hat{x}_{\min}$. Tests use nine typical {i.i.d.}, constrained {i.i.d.}, or constrained length $m + 1$ ordinal pattern (OP) power-law surrogates, have a nominal size 10\%, and use as test statistic (from left to right) the KS distance, mean, variance, maximum, conditional entropy of order one, and conditional entropy of order two. A solid gray line is drawn at the nominal size.
}
\label{fig:hyp_test_empirical_data_various_L}
\end{figure*}

\begin{figure*}[p]
\centering
\includegraphics[width=0.90\textwidth]{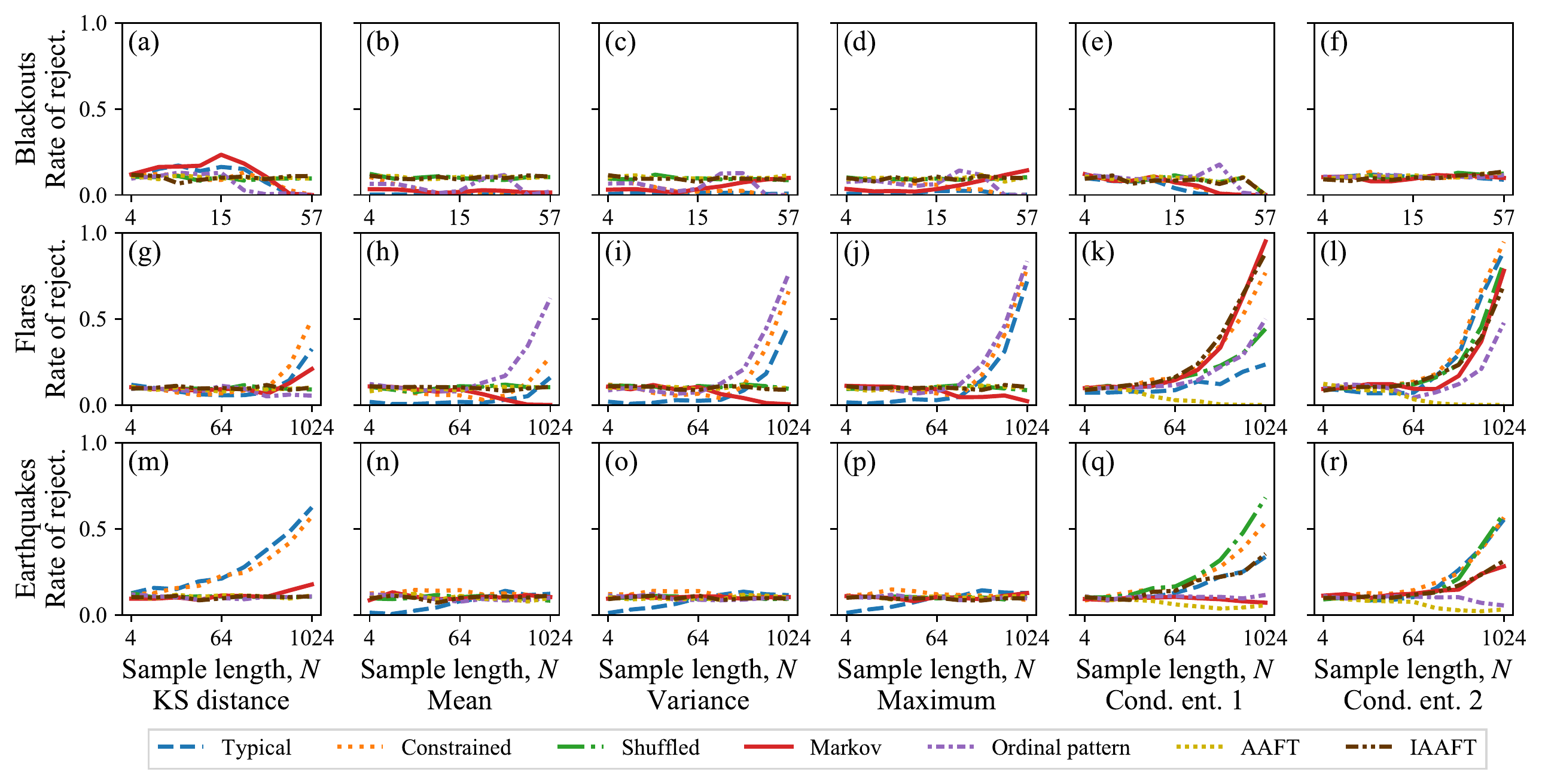}
\caption{{Incorporating correlation constraints impacts the rate of rejection of power-law hypotheses for empirical observations across a wide range of discriminating statistics.} 
Rates of rejection estimated from $1,000$ hypothesis tests of sequences of length $N$ of (a-f) number of customers affected by blackouts, (g-l) intensities of solar flares, and (m-r) energy released by earthquakes. Each input sequence $\xin$ considered begins at an independently and randomly chosen point in the full empirical sequence of values above the fitted lower cut-off $\hat{x}_{\min}$.  Tests use nine typical {i.i.d.} power-law, constrained {i.i.d.} power-law, shuffled, constrained Markov order one power-law, {constrained} length 16 ordinal pattern power-law, AAFT or IAAFT surrogates, have a nominal size 10\%, and use as test statistic (from left to right) the KS distance, mean, variance, maximum, conditional entropy of order one, and conditional entropy of order two. Markov power-law surrogates use the same Markov states as the original time series.
{When shuffle, AAFT or IAAFT surrogates (which only alter the order of a sequence, but not the frequencies of distinct values) are used together with KS distance, mean, variance or maximum (statistics independent of the order) the rate of rejection trivially remains close to the nominal size.}
}
\label{fig:hyp_test_empirical_data}
\end{figure*}

\begin{figure*}[p]
\centering
\includegraphics[width=0.50\textwidth]{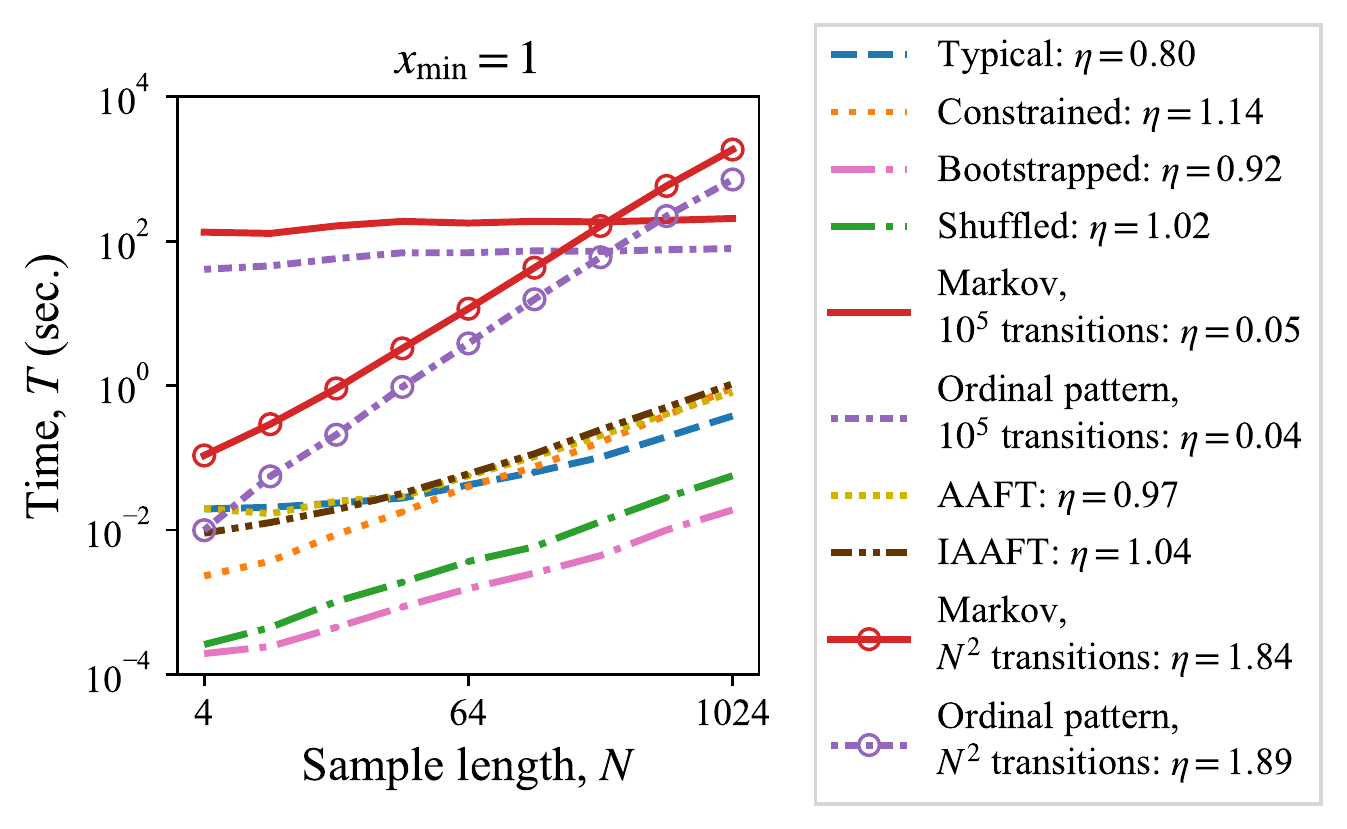}
\caption{
{Computational cost of surrogate methods as a function of sample size $N$. The computational cost of constrained (typical)} {i.i.d.} power-law surrogates scales {slightly more than (slightly less than) linearly} with {sample} length, while {the computational cost of constrained} Markov {order} and ordinal pattern power-law {surrogates} scales approximately linearly with the number of transitions. The mean over ten independent trials of the computational time $T$ required to produce 100 surrogates. Each input sequence $\xin$ arises independently from a single realisation of length $N$ of an i.i.d. power-law {process} with scale exponent $\alpha = 2.5$ {and lower cut-off $x_{\min} = 1$}. Typical {i.i.d.} power-law, constrained {i.i.d.} power-law, bootstrap, shuffle, Markov order one, ordinal pattern, amplitude adjusted Fourier transform (AAFT) and iterated amplitude adjusted Fourier transform (IAAFT) surrogates are considered. In the case of the constrained Markov order one power-law and ordinal pattern power-law surrogates, which use a Metropolis algorithm, we consider separately time required for $N^2$ and $10^5$ transitions per surrogate. The legend presents the exponent $\eta$ with which computational time $T$ scales with sample length $N$, according to {the model $T = K N^\eta$. The exponent $\eta$ is estimated using least squares regression of $\log T$ to $\log N$ for $N \geq 64$, using the average value of $T$ over the ten trials considered. The AAFT and IAAFT surrogate methods are MATLAB implementations called from within Python, which leads to communication costs which increase their apparent computational time.}
}
\label{fig:comp_time}
\end{figure*}


